\documentclass[A4paper,fleqn,usenatbib]{mnras}
\usepackage{threeparttable}
\usepackage{newtxtext,newtxmath}
\usepackage{soul}
\usepackage{hyperref}
\usepackage[T1]{fontenc}
\DeclareRobustCommand{\VAN}[3]{#2}
\let\VANthebibliography\thebibliography
\def\thebibliography{\DeclareRobustCommand{\VAN}[3]{##3}\VANthebibliography}
\def\simgt{\lower.5ex\hbox{$\; \buildrel > \over \sim \;$}}
\usepackage{ulem}
\usepackage{graphicx}	
\usepackage{amsmath}	
\usepackage{siunitx}
\usepackage{pdflscape}
\usepackage{tabularx}
\usepackage{longtable}
\usepackage{multirow}

\newcommand{\RNum}[1]{\uppercase\expandafter{\romannumeral #1\relax}}
\newcommand{\angstrom}{\textup{\AA}}

\defcitealias{zhu19}{Zhu19}
\defcitealias{wu13}{Wu13}

\title[X-ray enhancements of high-redshift RLQs]{The X-ray enhancements of radio-loud quasars at high redshift: New results at $z = 4\text{ -- }7$}

\author[Z. Zuo et al.]{
Zihao Zuo,$^{1}$
Shifu Zhu,$^{1,2,4,5}$\thanks{E-mail: SFZAstro@gmail.com}
W. N. Brandt,$^{1,2,3}$
Gordon P. Garmire,$^{6}$
F. Vito,$^{7}$
Jianfeng Wu,$^{8}$
\and
and Yongquan Xue$^{4,5}$
\\
$^{1}$Department of Astronomy \& Astrophysics, The Pennsylvania State University, University Park, PA 16802, USA\\
$^{2}$Institute for Gravitation and the Cosmos, The Pennsylvania State University, University Park, PA 16802, USA\\
$^{3}$Department of Physics, 104 Davey Lab, The Pennsylvania State University, University Park, PA 16802, USA\\
$^{4}$CAS Key Laboratory for Research in Galaxies and Cosmology, Department of Astronomy, University of Science and Technology of China, Hefei 230026, China\\
$^{5}$School of Astronomy and Space Sciences, University of Science and Technology of China, Hefei 230026, China\\
$^{6}$Huntingdon Institute for X-ray Astronomy, LLC, 10677 Franks Road, Huntingdon, PA 16652, USA\\
$^{7}$INAF – Osservatorio di Astrofisica e Scienza dello Spazio di Bologna, Via Gobetti 93/3, I-40129 Bologna, Italy\\
$^{8}$Department of Astronomy, Xiamen University, Xiamen, Fujian 361005, People’s Republic of China
}

\date{Accepted XXX. Received YYY; in original form ZZZ}

\pubyear{2023}

\begin{document}

\maketitle
\begin{abstract}
Highly radio-loud quasars (HRLQs; $\log R>2.5$) at $z\gtrsim 4$ show apparent
enhanced \hbox{X-ray} emission compared to matched HRLQs at lower redshifts,
perhaps due to a redshift-dependent fractional contribution to the \hbox{X-ray}
luminosity from inverse-Compton scattering of cosmic microwave background
photons (IC/CMB). Using new {\it Chandra\/} observations and archival
\hbox{X-ray} data, we investigate this phenomenon with an optically
flux-limited sample of 41 HRLQs at $z = 4$--5.5 all with sensitive \hbox{X-ray}
coverage, the largest sample utilized to date by a wide margin. \hbox{X-ray}
enhancements are assessed using \hbox{X-ray}-to-optical flux ratios and spectral
energy distributions. We confirm the presence of \hbox{X-ray} enhancements at a
\hbox{4.9--5.3$\sigma$} significance level, finding that the median factor
of enhancement is $\approx 1.8$ at our sample median redshift of $z\approx 4.4$.
Under a fractional IC/CMB model, the expected enhancement at lower redshifts is modest;
e.g., $\approx 4$\% at $z\approx 1.5$. We also investigate a sample of seven
radio-loud quasars (RLQs; $\log R>1$) at even higher redshifts of $z=5.6$--6.8,
using new and archival \hbox{X-ray} data. These RLQs also show evidence for
\hbox{X-ray} enhancements by a median factor of $\approx 2.7$ at a
\hbox{3.7--4.9$\sigma$} significance level. The \hbox{X-ray} spectral and
other properties of these $z=5.6$--6.8 RLQs, however, pose challenges for a
straightforward fractional IC/CMB interpretation of their enhancements. 
\end{abstract}


\section{Introduction}
Quasars are powered by the accretion process happening in the vicinity of supermassive black holes (SMBHs) located in the central regions of host galaxies. Quasars at high redshifts are of particular interest since they provide insights into SMBH growth and galaxy formation in the early universe. The discovery of significant samples of rare high-redshift quasars was made possible by large-scale sky surveys such as the Sloan Digital Sky Survey \citep[SDSS,][]{York2000}{}{}. Since Ly$\alpha$ emission is redshifted to the near-infrared (NIR) bands at $z\gtrsim5.5$, wide-field NIR surveys like the UKIRT Infrared Deep Sky Survey and the VISTA Hemisphere Survey \citep[e.g. ][]{UKIDSS,VHS} are also important. As of December 2022, there are 531 quasars with $z\geq 5.3$ in the literature \citep{Fan2022}. Based on the radio-loudness parameter $R=f_\mathrm{5GHz}/f_\mathrm{4400\angstrom}$, where $f_\mathrm{5GHz}$ and $f_\mathrm{4400\angstrom}$ are the flux densities at rest-frame \SI{5}{GHz} and \SI{4400}{\angstrom} \citep{Kellermann1989}, quasars can be divided into radio-quiet quasars (RQQs; $R\leq 10$) and radio-loud quasars (RLQs; $R> 10$). RLQs harbor strong relativistic jets \citep[e.g.][]{Padovani2017}. The fraction of RLQs remains $\sim10\%$ up to $z\sim 6$ \citep{banados2015}. It is well known that the quasar number density peaks at $2<z<3$ \citep[e.g.][]{Dunlop1990,Croom2004,Brown2006,Yang2016}{}{} and then rapidly declines with increasing redshift. In contrast to the strong redshift dependence of the quasar number density, previous studies have shown that the UV/optical and NIR spectral properties of quasars do not show significant evolution up to $z\sim6$ \citep[e.g.][]{Fan2004,Jiang2007,Shen2019}. Current observations suggest that AGNs and SMBH feeding modes do not show significant evolution after the first billion years of the Universe.

\par Almost all quasars are bright in the \mbox{X-ray} regime. The X-ray emission from RQQs is proposed largely to be generated by the thermal inverse-Compton (IC) process, where UV/optical photons provided by the accretion disk are upscattered by the hot electrons in a coronal structure \citep[e.g.][]{Haardt1993}. The $L_\mathrm{UV}$--$L_\mathrm{X}$ relation of RQQs also does not show significant evolution up to $z\sim6$ \citep[e.g.][]{steffen2006,Vito2019}{}{}. Previous studies found that the typical \mbox{X-ray} power-law photon index of $z>6$ RQQs is slightly steeper than that of RQQs at lower redshifts \citep[e.g.][]{Nanni2017,Vito2019,Wang2021}. A recent work conducted by \citet{Zappacosta2023} showed that the power-law photon index of a sample of hyperluminous (\mbox{$L_\mathrm{bol}\geq 10^{47}$ erg s$^{-1}$}) RQQs seems significantly steeper than that for RQQs at lower redshifts.
\par The nuclear \mbox{X-ray} emission of RLQs is stronger than that of RQQs with comparable UV/optical luminosities. Historically, this \mbox{X-ray} excess was believed to be produced by an additional jet-linked component \citep[e.g.][] {Zamorani1987,Worrall1987}{}{}. This idea has been challenged by \citet{zhu20,zhu21} and \citet{Timlin2021}, who proposed that the \mbox{X-ray} emission from most RLQs is still corona-dominated with stronger coronal X-ray emission caused by a corona-jet connection.

\par The \mbox{X-ray} luminosity contribution from the jet-linked component is much less important than previously expected for general RLQs. However, the many extended X-ray jets discovered by \textit{Chandra}\footnote{See, e.g., \url{https://hea-www.harvard.edu/XJET/}.}, often stretching over tens of kpc, demonstrate that some level of jet-linked X-ray emission exists. The radiation mechanisms of this jet-linked \mbox{X-ray} emission have been extensively discussed. One possible mechanism is seed photons being IC scattered by relativistic electrons accelerated in the jet. The seed photons of the IC process may come from the cosmic microwave background \citep[IC/CMB;][]{Tavecchio2000,Celotti2001}{}{}; synchrotron emission from the jet \citep[synchrotron self-Compton, or SSC;][]{Bloom1996}{}{}; or photons radiated from the accretion disc, broad-line region (BLR), or the dusty ``torus'' \citep[e.g.][]{Dermer1993,Sikora1994,Blazejowski2004}{}{}. The IC/CMB model was used to explain the \mbox{X-ray} knot emission of the kpc-scale jet of PKS 0637$-$752 \citep{Tavecchio2000,Chartas2000,Schwartz2000} and became a popular explanation for \mbox{X-ray} jets \citep[e.g.][]{Sambruna2004,Sambruna2006,Perlman2011,Marshall2018}{}{}. 

\par Despite the initial success of the IC/CMB model for extended \mbox{X-ray} jets, the lack of strong redshift
evolution of RLQ X-ray properties disfavors using IC/CMB to explain
the nuclear jet-linked X-ray emission \citep[e.g.][]{Bassett2004,Lopez2006,miller2011}. An \mbox{X-ray} jet-linked component that is dominated by IC/CMB emission would reproduce the $(1+z)^4$ dependence of the CMB energy density. \citet{miller2011} showed that the relative \mbox{X-ray} brightness of RLQs compared to RQQs with similar UV/optical luminosities does not show significant redshift dependence to $z\sim4$. However, \citet[][\citetalias{wu13} hereafter]{wu13} and \citet[][\citetalias{zhu19} hereafter]{zhu19} showed that the minority subset of highly radio-loud quasars (HRLQs, $\log R>2.5$) has an \mbox{X-ray} enhancement at $4<z<5.5$ compared to their matched low-$z$ counterparts. \citetalias{zhu19} formed a flux-limited sample of 24 HRLQs at $4<z<5.5$ (median $z=4.4$, $m_i\leq 20.26$) with complete \mbox{X-ray} coverage  and compared it with 311 HRLQs at $z<4$ (median $z=1.3$) with matched UV/optical luminosities. \citetalias{zhu19} found the nuclear \mbox{X-ray} emission of $4<z<5.5$ HRLQs is $\approx2$ times stronger than that for their low-$z$ counterparts at a \mbox{4--5 $\sigma$} confidence level. The \mbox{X-ray} enhancement factor of $\approx2$ is slightly smaller than the $\approx3$ factor found by \citetalias{wu13} but has a higher confidence level owing to the larger and improved sample. Although the $(1+z)^4$ dependence of the dominant \mbox{X-ray} emission has not been observed, contributions from the IC/CMB process are still expected. 

\par 
\citetalias{wu13} and \citetalias{zhu19} proposed a fractional IC/CMB model to explain the X-ray enhancements of high-redshift HRLQs. In this model, the IC/CMB process is only dominant on the scale of $\approx1\text{--}5$ kpc, where the CMB energy density starts to dominate those of other radiation fields and the magnetic field. In the inner region, the IC seed photons are dominated by the emission from the accretion process around the central SMBH, such as IR photons emitted by the torus \citep[e.g.][]{Ghisellini2009}{}{}, which does not show a strong redshift dependence. Here, \citet{Ghisellini2009} assumed a highly relativistic jet with a bulk Lorentz factor $\gamma=15$. On scales of $\gtrsim10$ kpc, the likely decelerated jet \citep[e.g.][]{Mullin2009,Breiding2017,Meyer2015, Meyer2016}{}{} can no longer boost CMB photons to X-ray energies. Indeed, if a jet with $\gamma\approx15$ at $\lesssim5$ kpc decelerates to $\gamma\approx1.5$ at a large distance, the X-ray emission produced by the IC/CMB process would decrease by a factor of $\approx100$. An alternative to the fractional IC/CMB model is that the enhanced star formation in high-$z$ host galaxies might create extra IR photons serving as seed photons for the IC process. It is also possible that high-$z$ jets are slower than those in the low-$z$ universe, leading to a larger fraction of beamed objects at $z>4$.

\par If the fractional IC/CMB model is correct, then RLQs at still higher redshifts of $z>5.5$ are expected to have strong IC/CMB-related \mbox{X-ray} emission due to the $(1+z)^4$ evolution of CMB energy density. Moreover, observations of several individual $z>5.5$ RLQs reveal inconsistent results regarding their X-ray enhancements \citep[e.g.][]{medvedev2020,Medvedev2021,connor2021,belladitta2020,Ighina2022}{}{}. Thus, it is important to study the role of the IC/CMB process in the early universe by investigating the \mbox{X-ray} enhancements of a sample of $z>5.5$ RLQs.

\par In this work, we investigate a sample of 41 HRLQs at $4<z<5.5$ and a sample of 7 RLQs at $z>5.5$. Our goals include the following: (1) further assess the existence of \mbox{X-ray} enhancements for $4<z<5.5$ HRLQs with an enlarged sample,  and (2) constrain the X-ray enhancements of RLQs at very high redshifts ($z>5.5$). In Section 2, we describe the selection of our $4<z<5.5$ HRLQ sample and the $z>5.5$ sample. We report our observations, analysis procedures, and \mbox{X-ray} properties of eight newly observed \textit{Chandra} Cycle 23 objects in Section 3. In Section 4, we present our main results and findings. The discussion and a summary are presented in Sections 5 and 6. Throughout the paper, we use short-hand nomenclature for object names in the text and Table \ref{tab:radioprop}. We adopt a flat $\Lambda$CDM cosmology with $H_0=$70.0 \mbox{km s$^{-1}$ Mpc $^{-1}$} and $\Omega_\mathrm{m}=0.3$.

\section{Sample Selection}
\label{sec:sample selection}
\subsection{The $4<z<5.5$ HRLQ sample}
\label{sec:HRLQ sample}
Our $4<z<5.5$ HRLQ sample was constructed based on the sources selected by \citetalias{wu13} and \citetalias{zhu19}. \citetalias{wu13} first selected four $z>4$ HRLQs from the SDSS Data Release 7 quasar catalog \citep[][]{DR7quasarDon} that covers a sky area of 9380 deg$^2$. They then found $24$ $z>4$ HRLQs within the sky area at $\delta>-40^\circ$ from NED.\footnote{\url{https://ned.ipac.caltech.edu/}} The sky area they chose was covered by the 1.4 GHz NRAO VLA Sky Survey \citep[NVSS;][]{NVSS}, which enabled them to determine the radio-loudness of each source. The flux densities of HRLQs ($\log R>2.5$) with $m_i\lesssim21$ at observed-frame 1.4 GHz should be above the NVSS detection limit ($\approx2.5$ mJy). The resulting sample contains 28 HRLQs (see Table 1 and Table 2 in \citetalias{wu13} for a full list).\footnote{The 28 HRLQs were revised to 26 by \citetalias{zhu19}. SDSS J003126.79+150739.5 and SDSS J123142.17+381658.9 were removed since they do not satisfy the HRLQ criterion ($\log R>2.5$) when using the rest-frame {2500 \angstrom} luminosity from \citet{shen2011}.} 

\par Based on the HRLQ sample constructed by \citetalias{wu13}, \citetalias{zhu19} found 17 more $4<z<5.5$ HRLQs.  \citetalias{zhu19} first matched the SDSS Data Release 14 quasar catalog \citep[][]{SDSSDR14Q} to the Faint Images of the Radio Sky at Twenty-centimeters survey \citep[FIRST;][]{FIRST} for observed-frame 1.4 GHz flux densities and found 16 HRLQs at $4<z<5.5$. FIRST should have detected all HRLQs with $m_i\lesssim21$ with a sensitivity of $\approx1$ mJy. \citetalias{zhu19} also searched NED following the method used by \citetalias{wu13} and selected another HRLQ at $z=4.067$, B3 0254+434 \citep[][]{Amirkhanyan2006}. 
\par
\begin{figure*}
\centerline{\includegraphics[scale=0.4]{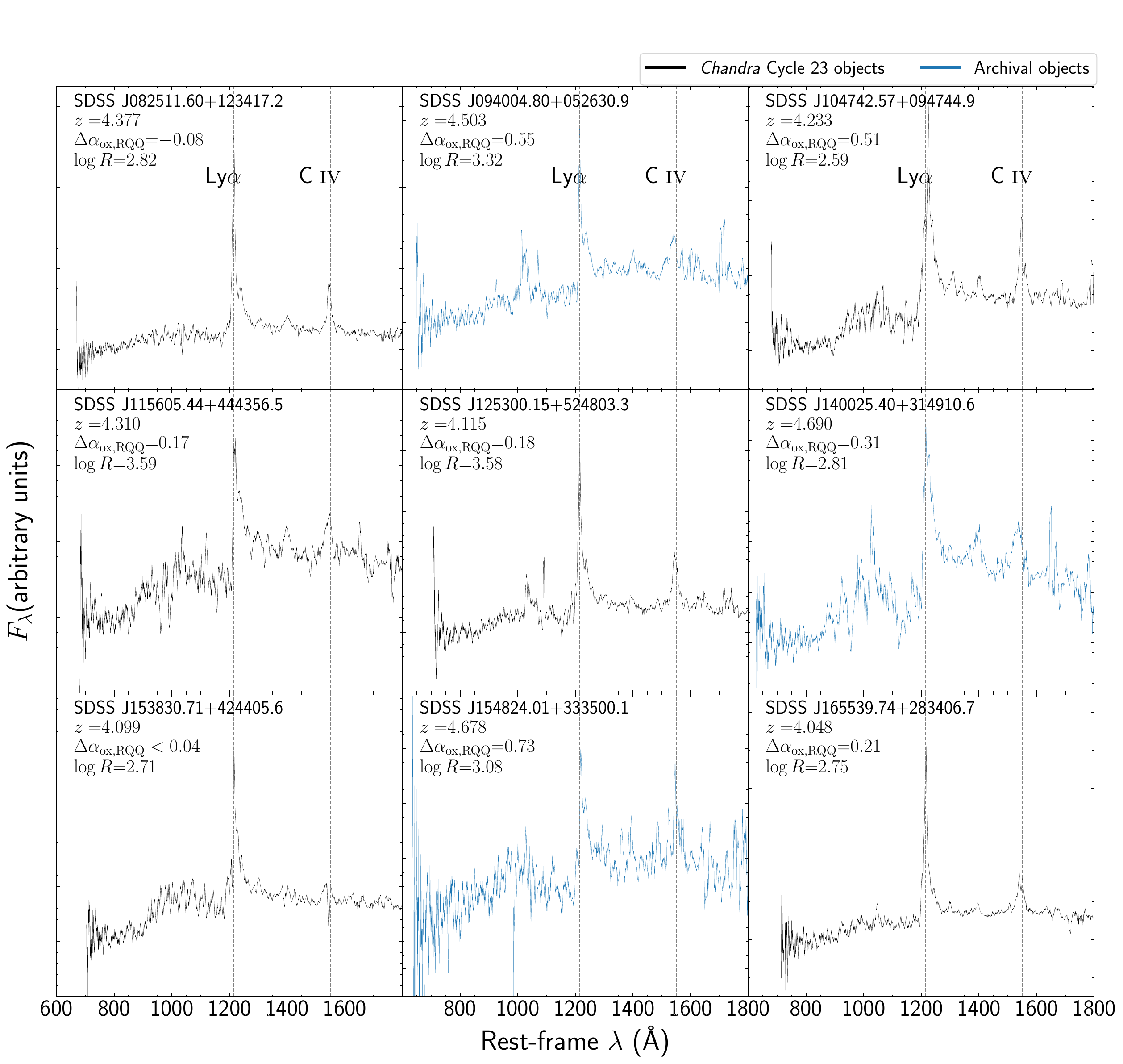}}
\caption{The rest-frame UV spectra of the new sources added to our sample, ordered by ascending RA. All spectra are taken from SDSS. The object name, redshift, $\Delta\alpha_{\mathrm{ox, RQQ}}$, and logarithm of the radio-loudness parameter are listed in each panel. The $x$-axis is the rest-frame wavelength in units of $\angstrom$. The $y$-axis is on a linear scale with arbitrary units. Each spectrum has been smoothed with a boxcar filter. Ly $\alpha$ ($\lambda=1215.24$ $\angstrom$) and C {\sc iv} ($\lambda=1549.48$ $\angstrom$) are labeled with the dotted vertical lines.}
\label{fig:spec}
\end{figure*}

 We combined the samples constructed by \citetalias{zhu19} and \citetalias{wu13}, with the two faintest sources ($m_i>22$) removed, to form a complete sample of $4<z<5.5$ HRLQs with 41 sources.  
32 out of 41 sources have been analyzed by \citetalias{wu13} and \citetalias{zhu19}. In this paper, we report the data analyses of the remaining nine objects using new and archival \textit{Chandra} observations. The rest-frame UV spectra of the nine newly-added HRLQs taken from SDSS are shown in \mbox{Fig. \ref{fig:spec}}. These objects generally show strong emission lines, indicating that their optical/UV continua are not strongly beamed. SDSS J1538+4244, showing an absorption feature on the blue wing of the C {\sc iv} emission line, was identified as a mini-Broad Absorption Line quasar \citep[mini-BAL; e.g.][]{Hall2002}{}{}. Since mini-BALs generally do not show weaker X-ray emission compared to typical quasars \citep[e.g.][]{Gibson2009,Wu2010}{}{}, we did not remove SDSS J1538+4244 from the sample.

\par We checked the VLA Sky Survey \citep[VLASS;][]{lacy2020}{}{} Epochs 1 and 2 Quick-look Catalogs \citep{Gordon2021} for potential source radio variability.\footnote{The catalogs were retrieved from \url{https://cirada.ca/vlasscatalogueql0}.} The majority of $4<z<5.5$ HRLQs do not show significant source variability during the 32 month time span between Epochs 1 and 2 of VLASS, corresponding to a rest-frame timescale of $\approx6.0$ months (calculated using the median redshift of $4<z<5.5$ HRLQs). All 41 HRLQs at $4<z<5.5$ are detected in both epochs. The flux densities at observed-frame 3 GHz of 40 objects fluctuate within 30\% between two epochs. The flux density of GB 1508+5714 obtained from VLASS Epoch 2 has doubled compared to that from \mbox{Epoch 1}. A $\approx30\%$ variability corresponds to a $\approx0.11$ fluctuation in $\log R$. The highly-radio-loud nature of $4<z<5.5$ HRLQs in our sample is thus not jeopardized by source variability.
\par
\begin{figure}
\centering
\includegraphics[width=\columnwidth]{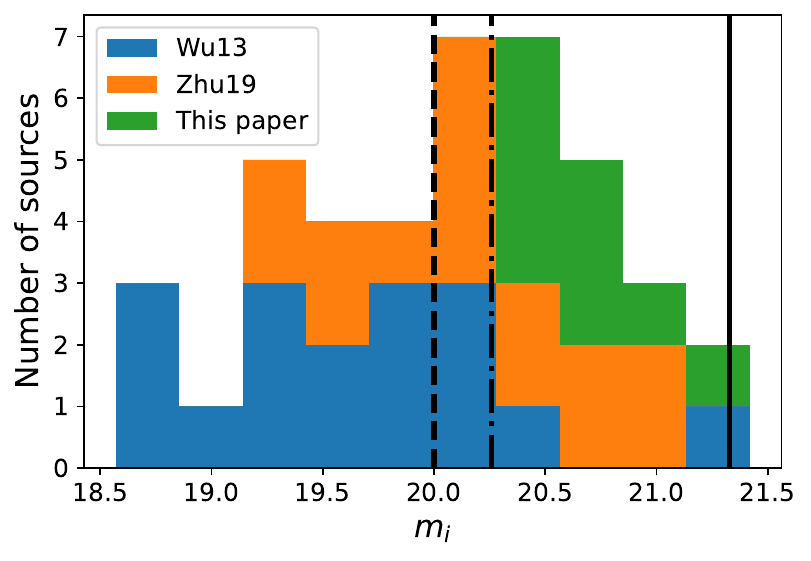}
\caption{The $m_i$ distribution for the $4<z<5.5$ HRLQ sample with 41 sources. The histogram is color-coded by the papers that provide \mbox{X-ray} analyses. The blue, orange, and green bars represent sources analyzed by \citetalias{wu13}, \citetalias{zhu19}, and this paper, respectively. The dashed, dash-dotted, and solid vertical lines are the flux limits applied in \citetalias{wu13}, \citetalias{zhu19}, and this paper, respectively. The flux-limited samples of \citetalias{wu13}, \citetalias{zhu19}, and this paper contain 12, 24, and 41 sources with sensitive \mbox{X-ray} coverage, respectively.}
\label{fig:sample_hist}
\end{figure}

 We plotted the $m_i$ distribution of the $4<z<5.5$ HRLQ sample in Fig. \ref{fig:sample_hist} and showed the joint efforts made by \citetalias{wu13}, \citetalias{zhu19}, and this paper to construct a high-redshift HRLQ sample with complete \mbox{X-ray} coverage over the past decade. The optically flux-limited sample utilized here is significantly larger than those of past work.

\subsection{The $z>5.5$ RLQ sample}
\label{sec:RLQ sample}
\par We collected $z>5.5$ RLQs from the literature \citep[e.g.][]{belladitta2020,banados2021,Ighina2021,Liu2021} and found 14 objects as of March 2021. Since the quasars selected by radio surveys tend to bias toward those with high radio luminosities, which may show more extreme properties, we removed three radio-selected quasars \citep[][]{McGreer2006,Zeimann2011,belladitta2020} and focused on optically selected objects. We also removed three optically selected putative RLQs since they were either not detected in deep follow-up radio observations \citep[][]{Liu2021} or the radio-loud nature depends on the assumed shape of the radio spectrum \citep[][]{banados2021}. Therefore, we found eight bona fide optical-selected RLQs at $z>5.5$, among which seven objects with $m_\mathrm{1450}<21.68$ have sensitive (new and archival) \textit{Chandra} or XMM-Newton coverage. The remaining object with $m_\mathrm{1450}=24.05$ still lacks sensitive X-ray data.

\par
\begin{figure*}
\centerline{\includegraphics[scale=0.6]{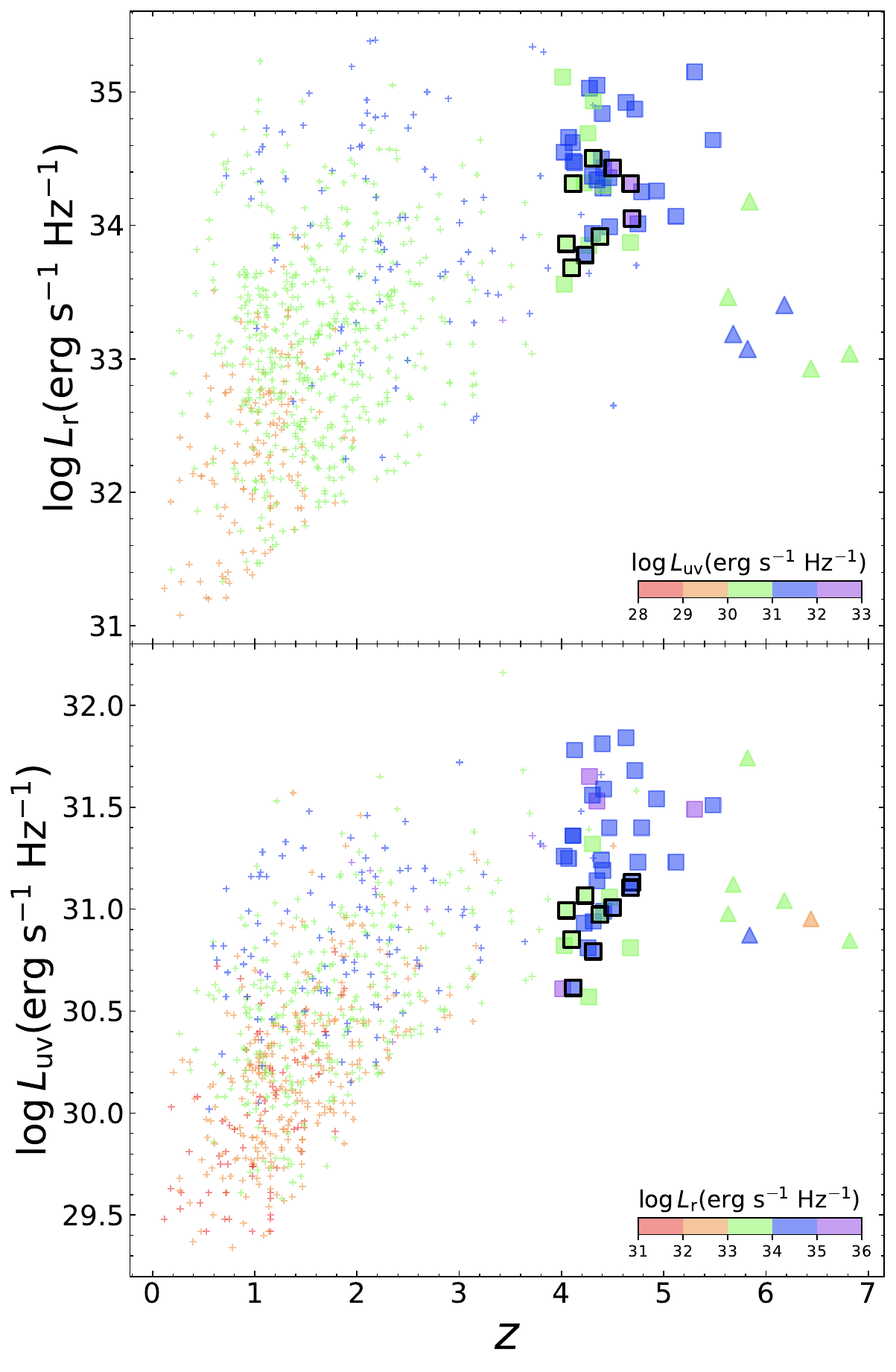}}
\caption{Radio (rest-frame 5 GHz, upper panel) and UV (rest-frame 2500 $\angstrom$, lower panel) monochromatic luminosities versus redshift. The upper and lower panels are color-coded by $\log L_\mathrm{uv}$ and $\log L_\mathrm{r}$, as indicated by the respective color bars. The squares represent 41 HRLQs at $4<z<5.5$ from \citetalias{zhu19}, \citetalias{wu13}, and this paper. Squares encircled by black boundaries are those newly added $4<z<5.5$ HRLQs. The triangles represent seven RLQs at $z>5.5$. The plus signs represent optically selected radio-loud quasars from \citet{zhu20}. Symbols in the upper and lower panels are color-coded based on their UV and radio monochromatic luminosities, respectively.}
\label{fig:logL}
\end{figure*}

 We plotted the radio and optical/UV luminosities against redshift for $4<z<5.5$ HRLQs, $z>5.5$ RLQs, and general RLQs from the full sample of \citet{zhu20} in Fig. \ref{fig:logL}. Compared to objects studied by \citetalias{wu13} and \citetalias{zhu19}, our objects are fainter in both the radio and optical/UV bands.
\par We repeated the VLASS radio variability analysis described in Section \ref{sec:HRLQ sample}. Three out of seven RLQs at $z>5.5$ were detected in both Epochs of VLASS and all of them show flux density fluctuations within 15\% on a rest-frame timescale of $\approx4.7$ months (calculated using the median redshift of $z>5.5$ RLQs). Only one object (VIK J2318$-$3113) was detected in one Epoch but not another. Observations at other radio frequencies further suggest that VIK J2318-3113 is probably a highly variable source. It is possible that such variability is extrinsic rather than intrinsic \citep{Ighina2022_VIK}. The typical $\approx15\%$ radio variability of $z>5.5$ RLQs translates to a $\approx0.06$ fluctuation in $\log R$, which does not jeopardize their radio-loud nature.

\section{observations and Data analyses}
In this section, we describe the data analyses of nine HRLQs at $4<z<5.5$ and seven RLQs at $z>5.5$ (see Table \ref{tab:obslog} for their observation logs).
\par Among the nine $4<z<5.5$ HRLQs, six have been awarded \textit{Chandra} Cycle 23 observation time and three are archival objects \citep[][]{snios2020}{}{} that are reanalyzed for consistency. Among the seven $z>5.5$ RLQs, two (PSO J055$-$00 and PSO J135+16) are \textit{Chandra} Cycle 23 objects. Two archival objects (VIK J2318$-$3113 and PSO J172+18) are analyzed and reported for the first time in this paper. Three additional objects with sensitive \mbox{X-ray} coverage reported in previous studies \citep{Brandt2002, connor2021, Medvedev2021} are reanalyzed for consistency.

\label{sec:data analyses}
\subsection{\textit{Chandra} data reduction and analysis}
\label{sec:chandra analyses}

All objects reported in Table \ref{tab:obslog} were observed with the Advanced CCD Imaging Spectrometer (ACIS) on board {\it Chandra} except for PSO J172+18. \footnote{CFHQS J1429+5447 has been observed by both \textit{Chandra} and \textit{XMM-Newton}. We used the most recent observation conducted by \textit{Chandra} in our analyses. However, the 0.5--2 keV flux derived from the \textit{Chandra} data is $\approx40\%$ fainter than that inferred from the \textit{XMM-Newton} data \citep{Medvedev2021}, suggesting the source is probably variable in \mbox{X-rays} on a rest-frame timescale of $\approx52$ days. The \textit{Chandra} observation also gave a flatter \mbox{X-ray} spectral index.} We utilized \textit{Chandra} Interactive Analysis of Observations ({\sc ciao}) version 4.14.0 to conduct X-ray data reduction. We first ran the \texttt{chandra\_repro} tool to reprocess and recalibrate the observation data with \texttt{CALDB} version 4.9.7. We did not see any object showing extended \mbox{X-ray} structures based on visual inspections. After that, we ran \texttt{fluximage} to generate exposure-corrected \mbox{X-ray} images in the full band \mbox{(0.5--8 keV)}, soft band \mbox{(0.5--2 keV)}, and hard band \mbox{(2--8 keV)}. Here we define the effective energy of each band to be the geometric mean of the energy limits. We then performed source detection by running \texttt{wavdetect} on the full-band image with a $10^{-6}$ significance threshold and wavelet scales $1,\sqrt{2},2,2\sqrt{2},$ and 4. All sources but SDSS J1538+4244 and VIK J2318$-$3113 were detected by \mbox{\texttt{wavdetect}}. We adopted \texttt{wavdetect} positions for the detected sources and the optical position for the undetected sources as source positions. We then applied the \texttt{deflare} procedure to background light curves to remove background flares and periods when the count rate was anomalously low. The \texttt{method} parameter is set to be \texttt{sigma} to adopt the \texttt{lc\_sigma\_clip} method with the default sigma value of 3, which iteratively deletes any point outside $\pm3\sigma$ about the mean until all points are in this range. When running \texttt{deflare}, regions of detected sources were masked out. Then, we filtered the event file with the good time interval (GTI) file produced by \texttt{deflare}. None of our observations was affected by flaring except for SDSS J1400+3149, whose exposure time has been reduced by 4\%.

\par We defined the source region to be a circle with a radius of 2.0 arcsec centered at the source position (corresponding to a $\approx94\%$ encircled-energy fraction, or EEF, of the PSF at 2 keV) and the background region to be an annulus centered at the source position, with inner and outer radii of 5.0 arcsec and 20.0 arcsec, respectively. We visually inspected the \mbox{X-ray} images to ensure that the background regions of reported objects are free of other X-ray sources. After that, we extracted the source and background counts of all three bands using \texttt{dmextract} with the source and background regions. We added the raw counts in the source and background regions of all five observations of VIK J2318$-$3113 together to get its raw source and background counts. We adopted the likelihood-ratio test developed by \citet{Li1983} for source detection. The null hypothesis is $s=0$, and the alternative hypothesis $s\neq 0$, where $s$ is the true source counts. Following \citet{Li1983}, we defined a likelihood ratio, $\lambda$, which is the ratio of the likelihoods that all \mbox{X-ray} counts are produced by the background over that there exists an actual \mbox{X-ray} source. The likelihood ratio, $\lambda$, satisfies Eq. (\ref{eq:1}), where $N_\mathrm{src}$ and $N_\mathrm{bkg}$ are source and background counts, and $B$ is background area to source area ratio:
\begin{multline}
    \label{eq:1}
    \ln\lambda=- \left( N_\mathrm{src}\ln\left((1+B)\frac{N_\mathrm{src}}{N_\mathrm{bkg}+N_\mathrm{src}}\right) \right.\\
     \left.+N_\mathrm{bkg}\ln\left(\frac{1+B}{B}\frac{N_\mathrm{bkg}}{N_\mathrm{bkg}+N_\mathrm{src}}\right) \right)
\end{multline}
The statistic $Z=\sqrt{-2\ln\lambda}$ is the Z-score of the source-detection test. That is, the likelihood-ratio test has a $Z$ $\sigma$ significance level. If the source counts are smaller than the estimated background counts, i.e., $N_\mathrm{src}<N_\mathrm{bkg}/B$, we flip the sign of $Z$ for this scenario. We rejected the null hypothesis if $Z\gtrsim2.576$, or equivalently $p$<0.01, and considered a source to be detected in the tested energy band. Here the $p$-value is calculated as $p=P( T >Z)$ where $T$ follows a standard normal distribution. The 0.01 threshold is suitable for source-detection tests with pre-specified positions.

\par We calculated $Z$ for all three bands of all our sources. For detected bands, we used a Bayesian method \texttt{aprates}\footnote{\url{https://cxc.cfa.harvard.edu/ciao/ahelp/aprates.html}.} \citep[e.g.][]{aprates}{}{} and set the \texttt{conf} parameter to 0.68 to calculate the net counts and their 1$\sigma$ credible intervals for all three bands, which are shown in Table \ref{tab:cnts}. For each undetected band, the upper bound of a 90 percent confidence interval of net counts is reported. Only a fraction of the source counts is included in the source aperture. We corrected the effects of nonideal PSF by setting the parameters \texttt{alpha} and \texttt{beta} of \texttt{aprates} to be the PSF fraction in the source aperture and background region, which are calculated using \texttt{src\_psffrac}.

\par We ran the \texttt{specextract} tool to extract source and background spectra. The \texttt{specextract} tool automatically generates ancillary response files (ARFs) and response matrix files (RMFs) except for SDSS J1538+4244, which does not have counts in the full band. We ran \texttt{mkacisrmf}, with the weight map provided by \texttt{specextract}, to create the weighted RMF and \texttt{mkarf} to create the ARF for SDSS J1538+4244. For VIK J2318$-$3113, we ran \texttt{combine\_spectra} to combine its spectra and response files from all five observations. We used the combined spectrum in the following analyses. 

\par We also calculated hardness ratios and their 1$\sigma$ confidence intervals using the Bayesian estimation method developed by \citet{park2006}. In this paper, hardness ratio $HR$ is defined as $\frac{H}{S}$, where $H$ and $S$ are net counts in the hard and soft bands, respectively. For each object detected in the soft band but undetected in the hard band, we reported an upper limit on the hardness ratio to be the hard count upper limit divided by soft counts. We then estimated the effective power-law photon index $\Gamma_\mathrm{X}$ using {\sc Sherpa}. For each object, we generated a series of Galactic-absorbed power-law spectra using the \texttt{xsphabs.abs1*powlaw1d.p1} model implemented in {\sc Sherpa}, which is the multiplication of an absorption model and a power-law model, with fixed Galactic neutral hydrogen column density ($N_\mathrm{H}$\footnote{\url{https://heasarc.gsfc.nasa.gov/cgi-bin/Tools/w3nh/w3nh.pl}}) and varying $\Gamma_\mathrm{X}$. We then used \texttt{calc\_model\_sum} to calculate the soft-band and hard-band counts of each model. With X-ray counts, we calculated the hardness ratio of each model and used interpolation to determine the desired $\Gamma_\mathrm{X}$ whose hardness ratio matches the observed value.

\par Our data reduction ends with running \texttt{calc\_energy\_flux} on the \texttt{xsphabs.abs1*powlaw1d.p1} model combining ARF and RMF files, $\Gamma_\mathrm{X}$, and $N_\mathrm{H}$ to derive the Galactic-absorption corrected soft-band flux. The model is normalized so that the result of running \texttt{calc\_model\_sum} on the soft-band equals the soft-band net counts or upper limit. 

\subsection{\textit{XMM-Newton} data reduction and analysis}
\label{sec: XMM reduction}

There are two archival X-ray observations of PSO J172+18 using EPIC 
onboard the {\it XMM-Newton} observatory.
We report the results of these 
observations (see Table \ref{tab:obslog} for the observation logs) for the first time.
We used the {\sc sas} package (v21.0.0) and the most recent Current Calibration Files 
to reprocess the {\it XMM-Newton} data. 
We used the {\sc epproc} command to produce calibrated event lists and filtered the resulting event lists using good time intervals created with 
single-event (i.e. pattern zero), and high-energy (10--12 keV) light curves with a criterion of ``RATE<=0.4''.
We extracted source spectra from a circular aperture with a 
radius of 15 arcsec centered at the optical position of PSO J172+18.
We also chose a circular source-free region with a radius of $65$ arcsec on 
the same CCD as the background region. We then extracted background spectra following the method we used to extract source spectra. The net counts, hardness ratio, and $\Gamma_\mathrm{X}$ are calculated following the steps described in Section \ref{sec:chandra analyses}. The results are reported in Table \ref{tab:cnts}.
The flux of PSO J172+18 appears only slightly above the sensitivity of these XMM-Newton observations. To achieve the best signal-to-noise ratio,
we only considered counts below 4.5 keV and defined soft, hard, and full bands as \mbox{0.3--1 keV}, \mbox{1--4.5 keV}, and \mbox{0.3--4.5 keV}, respectively.

\par
\begin{figure}
\centering
\includegraphics[width=0.48\textwidth, clip]{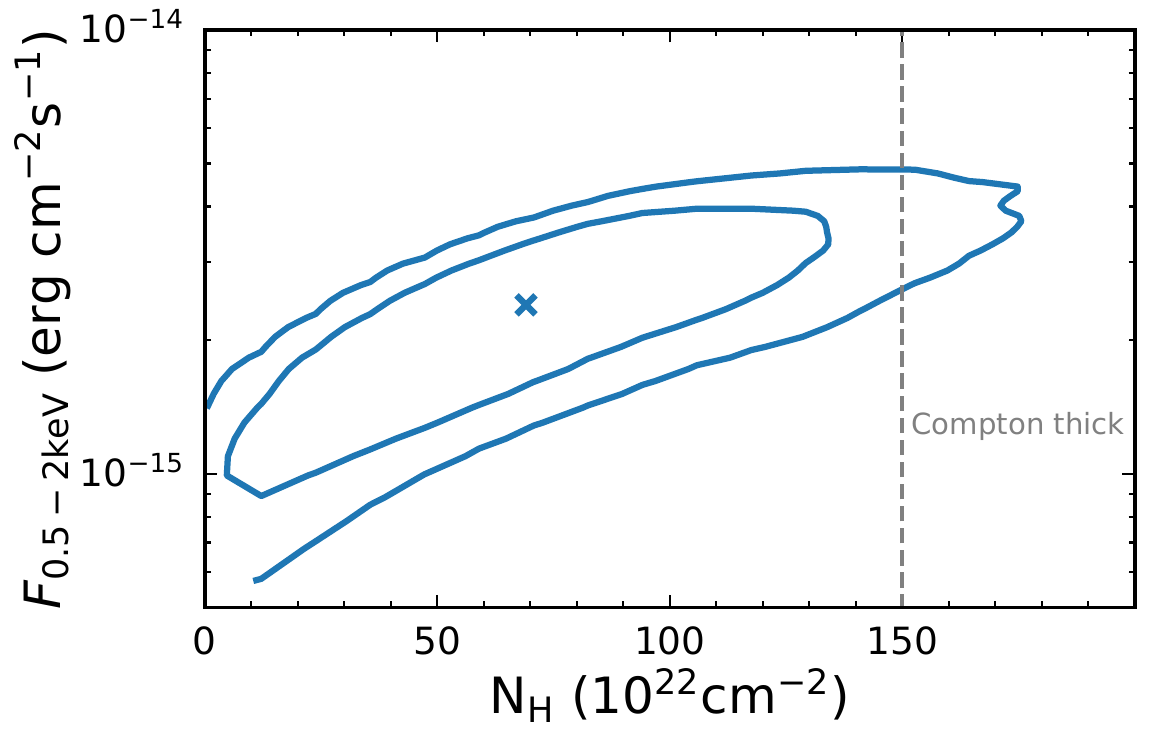}
\caption{The 68\% (inner) and 90\% (outer) confidence regions of the \mbox{0.5--2 keV} flux of the power-law continuum 
    and the column density of the intrinsic absorber for PSO J172+18. The photon index is fixed at $\Gamma_\mathrm{X}=2.0$. 
    The cross symbol indicates the best-fit model for $\Gamma_\mathrm{X}=2.0$, where the absorber is likely Compton thin with best-fit $N_\mathrm{H}=6.9\times10^{23}$ cm$^{-2}$.
    The observed X-ray flux is probably suppressed by a factor of $\approx 3$.
    }
\label{fig:confregion}
\end{figure}

 The X-ray emission of PSO J172+18 might suffer from additional absorption, given the flatness of the observed spectrum ($\Gamma_\mathrm{X}\approx1.17$; see Table \ref{tab:cnts}) and the faintness of the flux.\footnote{We examined the rest-frame optical/UV spectrum of PSO J172+18 provided by \citet{banados2021}. The optical/UV spectrum is similar to that of typical quasars, with prominent emission lines and a rest-frame UV power-law slope $\alpha_\nu=-0.48$. Thus, there is no apparent obscuration in the rest-frame optical/UV band.}
Since the \mbox{0.3--4.5 keV} band probes \mbox{2.3--35 keV} photons in the rest frame, the column density must be high if the putative absorber is intrinsic to PSO J172+18.
To assess the column density of the absorber and its impact on the X-ray fluxes,
we fit a power-law model with intrinsic photoelectric absorption in addition to the Galactic absorption.
The spectra from the two observations were fitted jointly.
Due to the limited number of net counts, the photon index and column density cannot be constrained simultaneously.
We thus fixed $\Gamma_\mathrm{X}$ to 2.0 \citep[e.g.][]{Vito2019,zhu21}{}{}, leaving the power-law normalization factor and the column density of the 
absorber free to vary. We performed a Markov chain 
Monte Carlo analysis to obtain a sample of 10$^4$ sets of free parameters using {\sc sherpa}.
The results are plotted in Fig.~\ref{fig:confregion}, where the normalization factor is replaced by absorption-corrected \mbox{0.5--2 keV} flux. If $\Gamma_\mathrm{X}=2.0$, the absorber is likely Compton thin with best-fit $N_\mathrm{H}=6.9\times10^{23}$ cm$^{-2}$.
We use the best-fit \mbox{0.5--2 keV} flux, \mbox{$2.40\times10^{-15}$ erg cm$^{-2}$ s$^{-1}$}, in later calculations.

\par Another possible explanation for the flat \mbox{X-ray} spectrum of PSO J172+18 is that it is a highly beamed object (or blazar). However, the steep radio spectrum \citep[$\alpha_\mathrm{r}=-1.31$,][]{banados2021}{}{} and the modest radio loudness ($\log R=2.07$) of PSO J172+18 disfavor this scenario.

\subsection{X-ray, optical/UV, and radio properties}
\label{sec: X-ray, optical/UV, and radio properties}
In this section, we elaborate on selected columns of Table \ref{tab:lowzresults} and Table \ref{tab:highzresults}, which report X-ray, optical/UV, and radio properties for our sample of $4<z<5.5$ HRLQs and $z>5.5$ RLQs. The captions of Table \ref{tab:lowzresults} and Table \ref{tab:highzresults} contain brief explanations of all columns.

\par The absolute $i$-band magnitude of the quasar is listed in column (3) of Table \ref{tab:lowzresults}. To be consistent with \citetalias{zhu19}, they are calculated from Galactic-extinction corrected  $m_i$ using equations (1) and (4) in Section 5 of \citet{Richards2006} and K-corrections from the same paper.
\par We adopted a new method to calculate the flux density observed at {$2500(1+z)$ \AA} for $z>5.5$ RLQs since \citet{Richards2006} did not provide K-corrections for $z>5.5$ quasars. We used the {\sc synphot} \citep{synphot} package to simulate the photometry of quasars observed in certain filters and adopted the composite quasar spectrum from \citet{vb2001} as the quasar spectrum template. We inspected the rest-frame optical/UV spectra of all seven $z>5.5$ RLQs and did not identify any objects with unusual continuum spectra. It is thus appropriate to use the composite spectrum given by \citet{vb2001} in our calculations. We then normalized the composite spectrum to the photometric data in the selected band and determined the flux density at {$2500(1+z)$ \angstrom} from the normalized composite spectrum.
Since rest-frame {$2500$ \angstrom} is in the observed-frame near-infrared (NIR) band at $z>5.5$, we used $H$-band magnitudes to calculate $f_{2500\angstrom}$ whenever possible. For objects without $H$-band observations, we used $J$-band (CFHQS J1429+5447 and PSO J352$-$15) or Pan-STARRS $y$-band (PSO J135+16) photometric data instead (see \ref{sec:sed}). The results are shown in column (9) of \hbox{Table \ref{tab:highzresults}}.

\par We calculated the radio spectral indices $\alpha_\mathrm{r}$ around observed-frame 1.4 GHz, with $f_\nu\propto\nu^{\alpha_\mathrm{r}}$. The results are shown in column (11) of Table \ref{tab:lowzresults} for HRLQs and column (11) of Table \ref{tab:highzresults} for $z>5.5$ RLQs. The 1.4 GHz flux densities were obtained from the FIRST or NVSS surveys. We also obtained flux densities at observed-frame 150 MHz, 326 MHz, 366 MHz, 3 GHz, and 5 GHz (See Section \ref{sec:sed}). We calculated $\alpha_\mathrm{low}$ and $\alpha_\mathrm{high}$, the radio spectral index between 1.4 GHz and a lower or higher frequency. We take the average value of $\alpha_\mathrm{low}$ and $\alpha_\mathrm{high}$ to be $\alpha_\mathrm{r}$.

\par The power-law spectral indices connecting rest-frame {2500 \AA} and 2 keV, defined by
\begin{equation}
    \alpha_\mathrm{ox}=\frac{\log(f_\mathrm{2keV}/f_\mathrm{2500\angstrom})}{\log(\nu_\mathrm{2keV}/\nu_\mathrm{2500\angstrom})},
\end{equation}
are shown in column (14) of Table \ref{tab:lowzresults} for HRLQs and column (14) of Table \ref{tab:highzresults} for $z>5.5$ RLQs.

\par We present differences between the measured $\alpha_\mathrm{ox}$ and the expected $\alpha_\mathrm{ox}$ value of RQQs ($\alpha_\mathrm{ox,RQQ}$) in column (15) of Table \ref{tab:lowzresults} for $4<z<5.5$ HRLQs and column (15) of Table \ref{tab:highzresults} for $z>5.5$ RLQs. More explicitly,
\begin{equation}
\label{eq: 3}
    \Delta\alpha_\mathrm{ox,RQQ}=\alpha_\mathrm{ox}-\alpha_\mathrm{ox,RQQ}.
\end{equation}
Here, $\alpha_\mathrm{ox,RQQ}$ is calculated using Eq. (3) of \citet{Just2007}:
\begin{equation}
\alpha_\mathrm{ox,RQQ}=-0.140\log L_\mathrm{2500\angstrom}+2.705.
\end{equation}

\par We also present differences between the measured $\alpha_\mathrm{ox}$ and the expected $\alpha_\mathrm{ox}$ of low-$z$ RLQs ($\alpha_\mathrm{ox,RLQ}$) in column (16) of Table \ref{tab:lowzresults} for $4<z<5.5$ HRLQs and column (16) of Table \ref{tab:highzresults} for $z>5.5$ RLQs. More explicitly,
\begin{equation}
    \Delta\alpha_\mathrm{ox,RLQ}=\alpha_\mathrm{ox}-\alpha_\mathrm{ox,RLQ}.
\end{equation}
Here, $\alpha_\mathrm{ox,RLQ}$ is calculated from the $L_\mathrm{2keV}$-$L_\mathrm{2500\angstrom}$-$L_\mathrm{5GHz}$ relation for the full RLQ sample given by model I in Table 4 of \citet{zhu20}, which is equivalent to
\begin{equation}
\label{eq:6}
    \alpha_\mathrm{ox,RLQ}= -0.203\log L_\mathrm{2500\angstrom} + 0.084\log L_\mathrm{5 GHz}+2.015.
\end{equation}

\section{Results}
\subsection{X-ray enhancements of $4<z<5.5$ HRLQs}
\label{sec:results-HRLQ}
\begin{figure*}
\centerline{\includegraphics[scale=0.6]{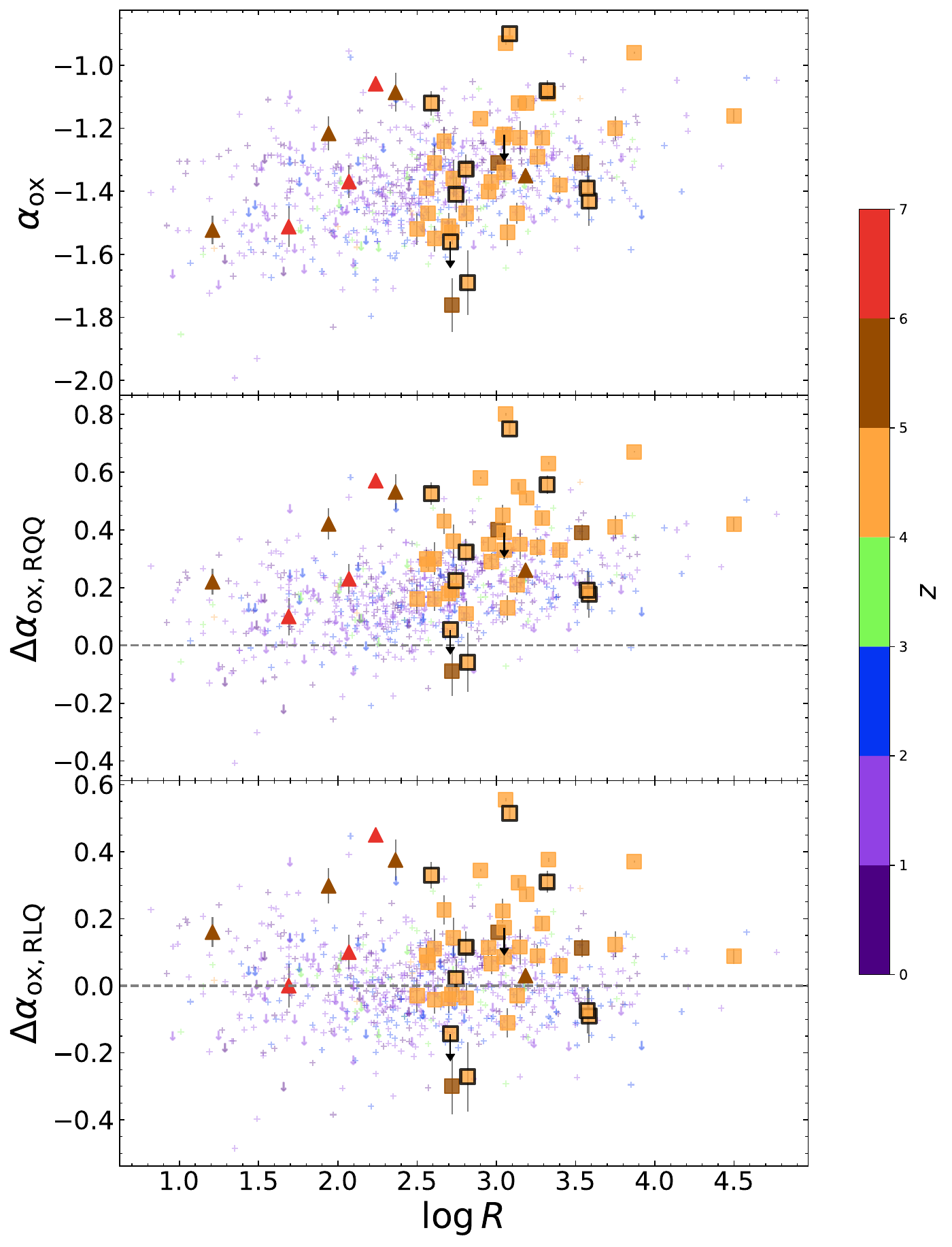}}
\caption{The $\alpha_{\mathrm{ox}}$ (upper panel), $\Delta\alpha_{\mathrm{ox,RQQ}}$ (middle panel), and $\Delta\alpha_{\mathrm{ox,RLQ}}$ (lower panel) parameters against logarithm of radio-loudness ($\log R$). The squares represent 41 HRLQs at $4<z<5.5$ from \citetalias{zhu19}, \citetalias{wu13}, and this paper. Squares encircled by black boundaries are those newly added $4<z<5.5$ HRLQs. The triangles represent seven RLQs at $z>5.5$.  The plus signs represent optically selected radio-loud quasars from \citet{zhu20} that are detected in \mbox{X-rays}. The downward arrows represent objects with only \mbox{X-ray} upper limits. All symbols are color-coded based on their redshifts. The data points in the lower two panels are not merely zero-point shifts. Compared to $\Delta\alpha_\mathrm{ox,RQQ}$, $\Delta\alpha_\mathrm{ox,RLQ}$ has an additional contribution from $\log L_\mathrm{5GHz}$ (see Eq. \ref{eq: 3}--\ref{eq:6}).}

\label{fig:logR_alpha}
\end{figure*}

To qualitatively assess the X-ray enhancements of $4<z<5.5$ HRLQs and $z>5.5$ RLQs, we plotted $\alpha_\mathrm{ox}$, $\Delta\alpha_\mathrm{ox,RQQ}$, and $\Delta\alpha_\mathrm{ox,RLQ}$ against $\log R$ for our full sample of $4<z<5.5$ HRLQs and $z>5.5$ RLQs in the upper, middle, and lower panels of Fig. \ref{fig:logR_alpha}, respectively. We also included optically selected radio-loud quasars from \citet{zhu20} in Fig. \ref{fig:logR_alpha} for comparison. The full sample of \citet{zhu20} has 729 RLQs, among which 657 (90.1\%) objects have X-ray detections. The redshift of the comparison sample has a median of 1.5 and an interquartile range (IQR) of 1.0. We plotted horizontal dashed lines with $\Delta\alpha_\mathrm{ox,RQQ}=0$ and $\Delta\alpha_\mathrm{ox,RLQ}=0$ in the middle and lower panels of Fig. \ref{fig:logR_alpha}. The positions of $4<z<5.5$ HRLQs (stars, squares, and diamonds) in the middle and lower panels of Fig. \ref{fig:logR_alpha} are generally higher than the low-$z$ comparison RLQs, implying enhanced X-ray emission. More specifically, 39 out of 41 HRLQs at $4<z<5.5$ have positive $\Delta\alpha_\mathrm{ox,RQQ}$, and 29 have positive $\Delta\alpha_\mathrm{ox,RLQ}$.
\par
\begin{figure*}
\centerline{\includegraphics[scale=0.6]{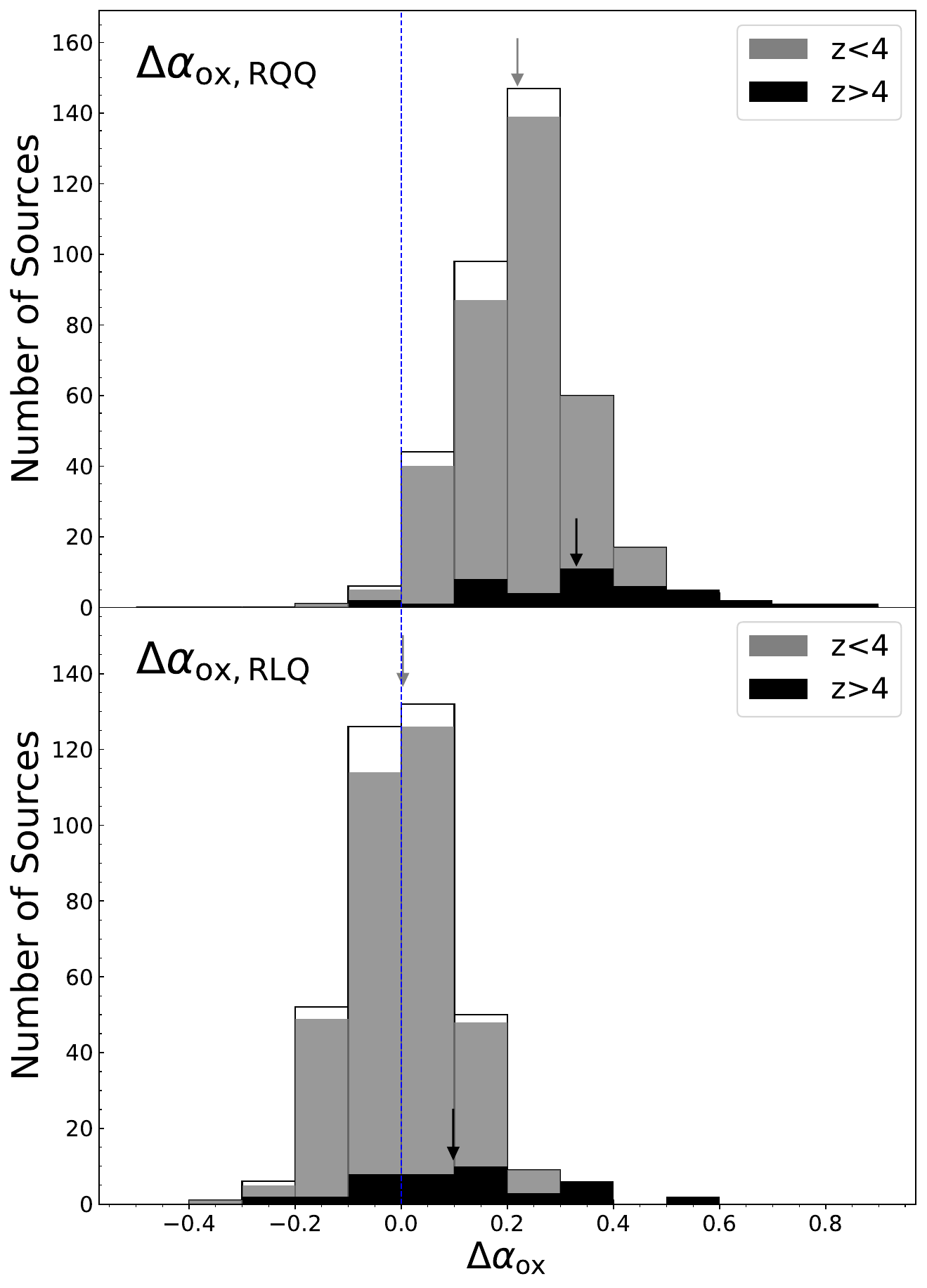}}
\caption{The grey and open bars are histograms of $\Delta\alpha_\mathrm{ox,RQQ}$ (upper panel) and $\Delta\alpha_\mathrm{ox,RLQ}$ for the optically selected RLQs from \citet{zhu20} with $z<4,m_i<21.33,$ and $\log R>2.5$, where grey bars are X-ray detected objects and open bars are X-ray undetected objects. The black bars are sources from the combined sample of \citetalias{wu13}, \citetalias{zhu19}, and this work. The downward arrows indicate the medians of $\Delta\alpha_\mathrm{ox}$ distributions for HRLQs at $z<4$ and $z>4$, respectively.}
\label{fig:hist}
\end{figure*}

 We then applied three conditions ($z<4$, $m_i<21.33$, $\log R>2.5$) to the full sample of \citet{zhu20} to form a sample of $z<4$ HRLQs that are comparable to $z>4$ HRLQs in the optical/UV and radio bands. Here we adopted an upper limit of 21.33 for $m_i$ since all objects in our $4<z<5.5$ HRLQ sample are optically brighter than $m_i=21.33$. We plotted the distributions of $\Delta\alpha_\mathrm{ox,RQQ}$ and $\Delta\alpha_\mathrm{ox,RLQ}$ for the low-$z$ HRLQ comparison sample and the $4<z<5.5$ HRLQ sample in Fig. \ref{fig:hist} for quantitative analysis. The low-$z$ HRLQ sample contains 377 objects with a median redshift of 1.5 and an IQR of 1.0. The logarithm of the radio-loudness parameter ($\log R$) of the comparison sample has a median of 3.0 and an IQR of 0.6. 
\par
\begin{figure*}
\centerline{\includegraphics[scale=0.5]{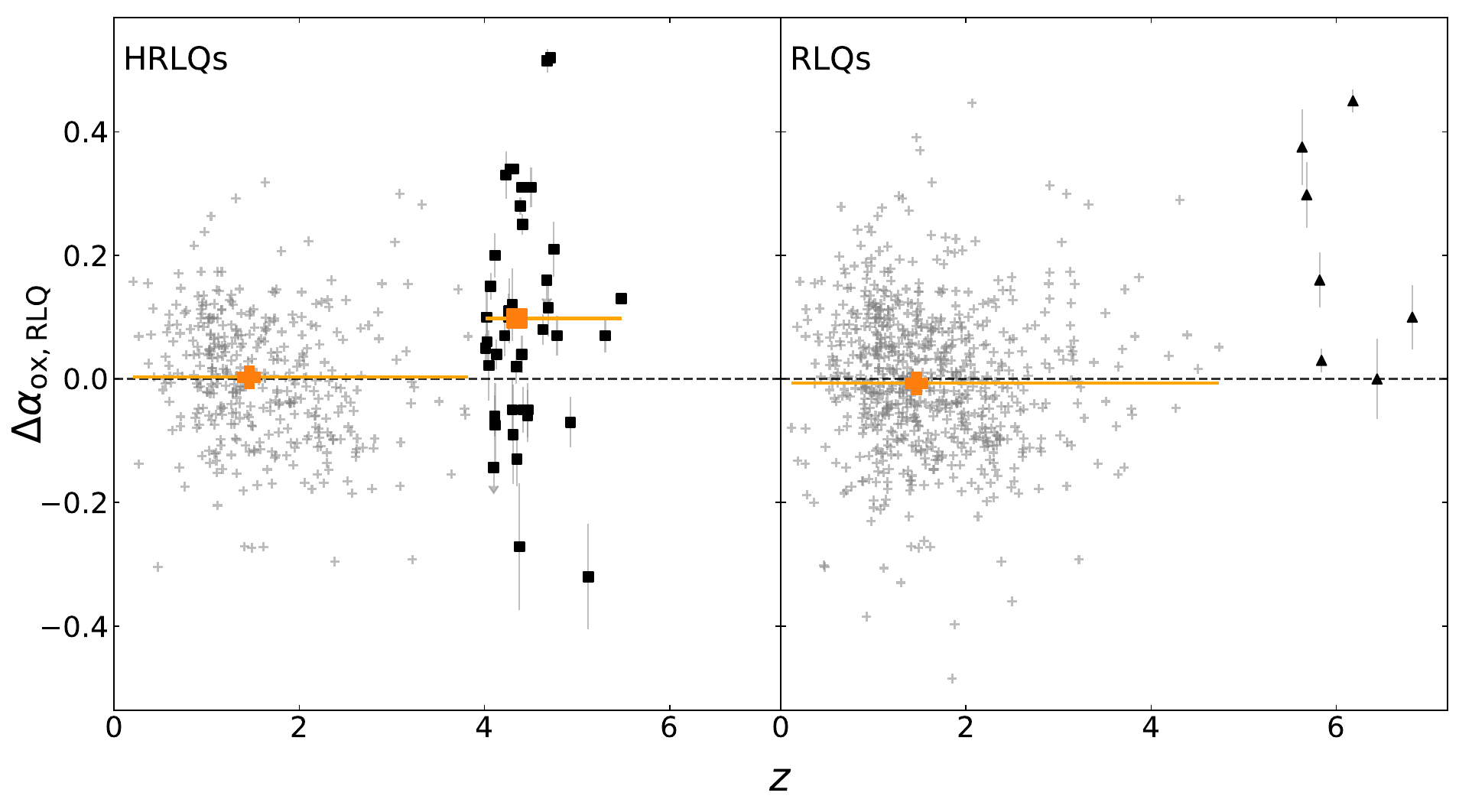}}
\caption{$\Delta\alpha_\mathrm{ox,RLQ}$ versus redshift for HRLQs (left panel) and RLQs (right panel). \textbf{Left:} The grey plus signs represent the low-redshift HRLQ comparison sample. The squares are 41 HRLQs at $4<z<5.5$. The orange plus sign represents the median $\Delta\alpha_\mathrm{ox,RLQ}$ of the low-redshift HRLQ comparison sample. The orange square represents the median $\Delta\alpha_\mathrm{ox,RLQ}$ of the $4<z<5.5$ HRLQ sample. \textbf{Right:} The grey plus signs represent the low-redshift RLQ comparison sample. The triangles are seven RLQs at $z>5.5$. The orange plus sign represents the median $\Delta\alpha_\mathrm{ox,RLQ}$ of the low-redshift RLQ comparison sample. The orange triangle represents the median $\Delta\alpha_\mathrm{ox,RLQ}$ of the $z>5.5$ RLQ sample. For both panels, orange horizontal error bars represent redshift ranges of relevant datasets. Orange vertical error bars represent 1$\sigma$ uncertainties of the medians estimated using the bootstrap.}
\label{fig:daox_z}
\end{figure*}

 Since our data include upper limits, we adopted survival analysis methods implemented in the {\sc python} package {\sc lifelines} \citep{lifelines} in the following calculations. We used the Kaplan-Meier estimator to estimate the cumulative distribution functions of all datasets. We then calculated the medians of all distributions from the cumulative distribution functions. We used bootstrap (1000 times) to estimate the $1\sigma$ uncertainties of the medians we calculated in the previous step. The medians of $\Delta\alpha_\mathrm{ox,RQQ}$ for low-$z$ and high-$z$ HRLQs are $0.219_{-0.004}^{+0.005}$ and $0.334_{-0.034}^{+0.016}$, respectively. The medians of $\Delta\alpha_\mathrm{ox,RLQ}$ for low-$z$ and high-$z$ HRLQs are $0.003_{-0.004}^{+0.005}$ and $0.098^{+0.016}_{-0.011}$, respectively. The difference between the medians of the $\Delta\alpha_\mathrm{ox,RLQ}$ distributions at $4<z<5.5$ and $z<4$ is $0.095^{+0.017}_{-0.012}$. Our results imply that the  X-ray emission of high-$z$ HRLQs is $1.77^{+0.19}_{-0.12}$ times stronger than that of their \mbox{low-$z$} counterparts. After that, we used the Peto-Peto test implemented in the {\sc lifelines} package to test the significance of the difference between $\Delta\alpha_\mathrm{ox}$ distributions of low-$z$ and high-$z$ HRLQs. The statistical test shows a $5.33\sigma$ ($p=5.01\times10^{-8}$) difference for $\Delta\alpha_\mathrm{ox,RQQ}$ distributions and a $4.90\sigma$ ($p=4.74\times10^{-7}$) difference for $\Delta\alpha_\mathrm{ox,RLQ}$ distributions. To illustrate the redshift evolution of $\Delta\alpha_\mathrm{ox,RLQ}$, we plotted $\Delta\alpha_\mathrm{ox,RLQ}$ versus redshift for the low redshift HRLQ comparison sample (grey plus signs) and the $4<z<5.5$ HRLQ sample (black squares) in the left panel of Fig. \ref{fig:daox_z}.
\par We performed a two-sample Kolmogorov-Smirnov (KS) test on the $\log R$ distributions of the $4<z<5.5$ HRLQ sample and the $z<4$ HRLQ comparison sample. With the null hypothesis that the $\log R$ distributions from the two samples are the same, the KS test gave a $p$-value of 0.95. We also performed a similar KS test on the radio flux distributions and found a $p$-value of 0.30. The medians ($\sigma_\mathrm{NMAD}$\footnote{$\sigma_\mathrm{NMAD}=1.4826\times \text{median absolute deviation (MAD)}$. This is a robust estimator of standard deviation.
}) of the $\log R$ distributions of $4<z<5.5$ HRLQs and $z<4$ HRLQs are $3.04^{+0.02}_{-0.09}(0.43)$ and $2.98^{+0.04}_{-0.03}(0.44)$, respectively. Similarly, for the log radio flux distributions of the high-$z$ and low-$z$ HRLQs, the medians ($\sigma_\mathrm{NMAD}$) are $-24.16^{+0.03}_{-0.03}(0.65)$ and $-24.19^{+0.07}_{-0.01}(0.52)$, respectively. Therefore, both the $\log R$ values and radio fluxes are consistent among the two samples. Thus, we do not need an explicit cut on $\log R$ or radio fluxes.
\par The results from \citetalias{zhu19} indicate that the X-ray enhancement of high-$z$ HRLQs is a factor of $1.9^{+0.5}_{-0.4}$ at a $4.07\sigma$ significance level considering the differences between $\Delta\alpha_\mathrm{ox,RLQ}$ distributions of \mbox{low-$z$} and high-$z$ HRLQs. Compared to \citetalias{zhu19}, our results indicate a slightly lower X-ray enhancement factor (though statistically consistent within uncertainties) at a higher significance level. The X-ray enhancement factor for our results is smaller because only five out of nine newly added $4<z<5.5$ HRLQs have positive $\Delta\alpha_\mathrm{ox,RLQ}$. However, since the sample size has been improved from 24 (the flux-limited sample in \citetalias{zhu19}) to 41, the significance level of the X-ray enhancement has been improved.

\subsection{X-ray enhancements of $z>5.5$ RLQs}
\label{sec:results-RLQ}
Referring to the middle and lower panels of Fig. \ref{fig:logR_alpha}, the positions of the seven $z>5.5$ RLQs are generally higher than the comparison sample. All seven objects have positive $\Delta\alpha_\mathrm{ox,RQQ}$, and non-negative $\Delta\alpha_\mathrm{ox,RLQ}$. Five of them have $\Delta\alpha_\mathrm{ox,RLQ}\geq 0.10$, and three stand out with $\Delta\alpha_\mathrm{ox,RLQ}\geq 0.30$.

\par We used the sample of optically selected RLQs from \citet{zhu20} with $m_i\approx m_\mathrm{1450}<21.68$ as the comparison sample. Here the flux limit is the $m_{1450}$ of the faintest $z>5.5$ RLQ. The \mbox{low-redshift} comparison sample of $z>5.5$ RLQs differs from that of $4<z<5.5$ by removing the $\log R>2.5$ and $z<4$ conditions and changing the flux limit. We then followed the methods used in Section \ref{sec:results-HRLQ} to calculate the medians of the $\Delta\alpha_\mathrm{ox}$ distributions and test the significance levels of the differences between the $\Delta\alpha_\mathrm{ox}$ distributions for RLQs at $z\lesssim5$ and $z>5.5$. The medians of $\Delta\alpha_\mathrm{ox,RQQ}$ at $z\lesssim5$ and $z>5.5$ are $0.173^{+0.007}_{-0.006}$ and $0.269^{+0.167}_{-0.025}$, respectively. The medians of $\Delta\alpha_\mathrm{ox,RLQ}$ at $z\lesssim5$ and $z>5.5$ are $-0.007_{-0.006}^{+0.005}$ and $0.160^{+0.140}_{-0.060}$, respectively. The difference between the medians of the $\Delta\alpha_\mathrm{ox,RLQ}$ distributions at $z>5.5$ and $z\lesssim5$ is $0.167^{+0.140}_{-0.060}$. This corresponds to an X-ray enhancement factor of $2.72^{+3.58}_{-1.47}$. The large uncertainties are due to the small sample size at $z>5.5$. The statistical test shows a $3.65\sigma$ \mbox{($p=1.30\times10^{-4}$)} difference for the $\Delta\alpha_\mathrm{ox,RQQ}$ distributions and a $4.91\sigma$ \mbox{($p=4.61\times10^{-7}$)} difference for the $\Delta\alpha_\mathrm{ox,RLQ}$ distributions. The results are not significantly affected if the potentially heavily absorbed source, PSO J172+18, is removed. Following Section \ref{sec:results-HRLQ}, we plotted $\Delta\alpha_\mathrm{ox,RLQ}$ versus redshift for the low redshift RLQ comparison sample (grey plus signs) and the $z>5.5$ RLQ sample (black triangles) in the right panel of Fig. \ref{fig:daox_z}.
\par A KS test on the $\log R$ distributions of the $z>5.5$ RLQ sample and its low-redshift comparison sample gives a $p$-value of 0.07. A similar KS test on radio flux distributions gives a $p$-value of 0.003. The KS tests suggest that the $\log R$ values, as well as the radio fluxes, are marginally different among the two samples. If such differences do exist, the low-redshift comparison sample could only bias toward higher $\log R$ (see Fig. \ref{fig:logR_alpha}) and radio fluxes. Since objects with lower $\log R$ and radio fluxes are expected to have less jet-linked \mbox{X-ray} emission, we conservatively did not apply an upper limit on $\log R$ or radio fluxes for the comparison sample.

\subsection{Spectral energy distributions}
\label{sec:sed}
To compare the broadband spectral energy distributions (SEDs) of HRLQs at $4<z<5.5$ and RLQs at $z>5.5$ with their low-$z$ counterparts, we constructed the SEDs for the nine newly added $4<z<5.5$ HRLQs and the seven $z>5.5$ RLQs. The SEDs of the remaining $4<z<5.5$ HRLQs can be found in \citetalias{wu13} and \citetalias{zhu19}. This comparison serves as an additional test of \mbox{X-ray} enhancements of $4<z<5.5$ HRLQs and $z>5.5$ RLQs. The multiwavelength photometric data were collected from the following sources:
\begin{enumerate}
    \item Radio: See Table \ref{tab:radioprop} and its footnotes for detailed information about the available radio frequencies for each source.
    
    \item Submillimeter: We included submillimeter photometry for $z>5.5$ RLQs. The flux densities at \SI{1.2}{\milli\meter} of CFHQS J1429+5447 and PSO J352$-$15 are taken from \citet{Khusanova22}. The flux densities at \SI{1.2}{\milli\meter} of VIK J2318$-$3113 and SDSS J0836+0054 are taken from \citet{Venemans2020} and \citet{petric2003}, respectively. The flux densities at \SI{850}{\micro\metre} of PSO J055$-$00 and PSO J135+16 are taken from \citet{Li2020}. PSO J055$-$00 is a non-detection; we took its 2$\sigma$ upper limit. 
    \item Mid-infrared: The mid-infrared flux densities are taken from the WISE all-sky catalog \citep{WISE} except for SDSS J1548+3335, PSO J055$-$00, VIK J2318$-$3113, PSO J352$-$15, and PSO J172+18, which are non-detections in the WISE all-sky catalog. We retrieved the $W1$ and $W2$ magnitudes of the first four sources from the CatWISE 2020 catalog \citep{CatWISE}. We used the WISE all-sky catalog whenever possible since the CatWISE catalog did not provide $W3$ and $W4$ magnitudes. Following \citet{banados2021}, we took the $W1$ and $W2$ magnitudes of PSO J172+18 from the DESI Legacy Imaging Surveys \citep[DECaLS;][]{Dey2019}{}{}.
    \item Near-infrared: The flux densities in near-infrared bands are taken from the UKIRT Hemisphere Survey \citep[UHS;][]{UHS}, the VISTA Kilo-degree Infrared Galaxy Survey \citep[VIKING;][]{Edge2013}{}{}, the VISTA Hemisphere Survey \citep[VHS;][]{VHS}, or dedicated follow-up observations. SDSS J1253+5248 and SDSS J1655+2834 only have $J$-band detections from UHS. PSO J055$-$00 has $J$, $H$, and $K$ band detections from VHS. SDSS J0836+0054 and VIK J2318$-$3113 have $J$, $H$, and $K$ band detections from VIKING. PSO J352$-$15 has $Y$ and $J$ band detections from VHS. CFHQS J1429+5447 has $J$-band detection from the Canada–France High-$z$ Quasar Survey \citep[CFHQS;][]{CFHQS2010}{}{}. SDSS J1548+3335 was covered by UHS but not detected in any band. Due to our particular interest in this source, we set an upper limit for its $J$-band flux by taking the maximum $J$-band magnitude among all detections within 1 arcmin of it. We took $J$, $H$, and $K$ band magnitudes for PSO J172+18 from follow-up observations reported by \citet{banados2021}.
    \item Optical: Considering the effects of the Ly$\alpha$ forest, we took SDSS $r$, $i$, and $z$ magnitudes and Pan-STARRS $y$ magnitudes for $4<z<5.5$ objects. For objects with larger redshift ($z>5.5$), we took Pan-STARRS $z$ and $y$ magnitudes. VIK J2318$-$3113 is not covered by Pan-STARRS.
    \item X-ray: The flux densities at observed-frame 2 keV of SDSS J0836+0054, CFHQS J1429+5447, and PSO J352$-$15 were retrieved from \citet{Brandt2002}, \citet{Medvedev2021}, and \citet{connor2021}, respectively. The observed-frame 2 keV flux densities of all other sources are from this paper.

\end{enumerate}
\par
\begin{figure*}
\centerline{\includegraphics[scale=0.5]{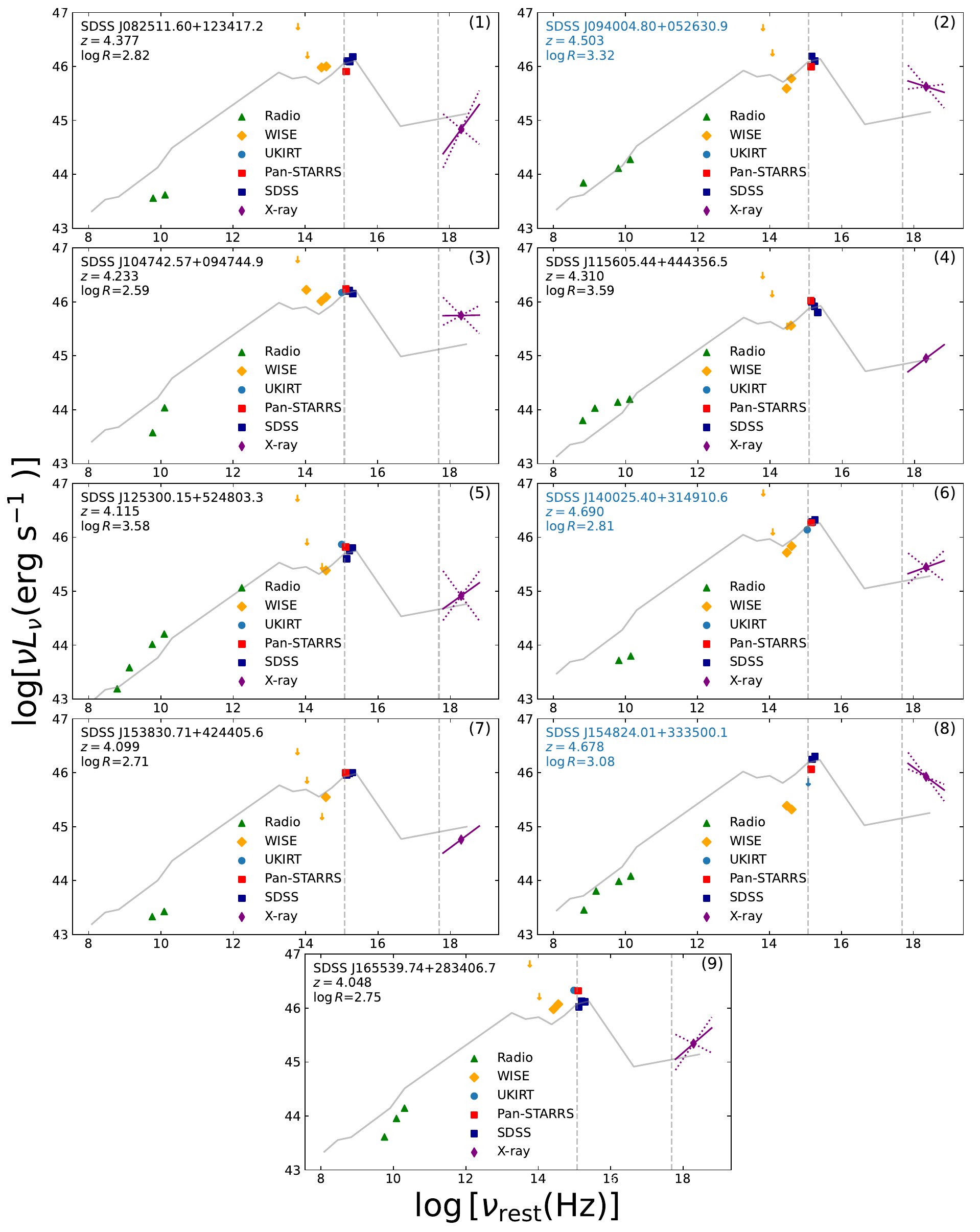}}
\caption{SEDs of the nine newly added $4<z<5.5$ HRLQs, ordered by ascending RA. Downward arrows represent upper limits. Objects with black text are \textit{Chandra} Cycle 23 objects. Those with blue text are archival objects. The purple diamonds are located at observed-frame 2 keV. The solid purple lines are X-ray power-law spectra around observed-frame 2 keV derived from photon indices reported in Table 3. Their uncertainties are represented by the dotted purple lines. The solid grey curves are composite SEDs for the 10 HRLQs at $z<1.4$ from \citet{Shang2011} with comparable optical luminosity and radio loudness, scaled to the flux density at rest-frame 2500 \AA. The dotted grey vertical lines are at rest-frame 2500 {\AA} and 2 keV. The name, redshift, and $\log R$ of each object are shown at the upper left of each panel.   }
\label{fig:sedlowz}
\end{figure*}

\begin{figure*}
\centerline{\includegraphics[scale=0.5]{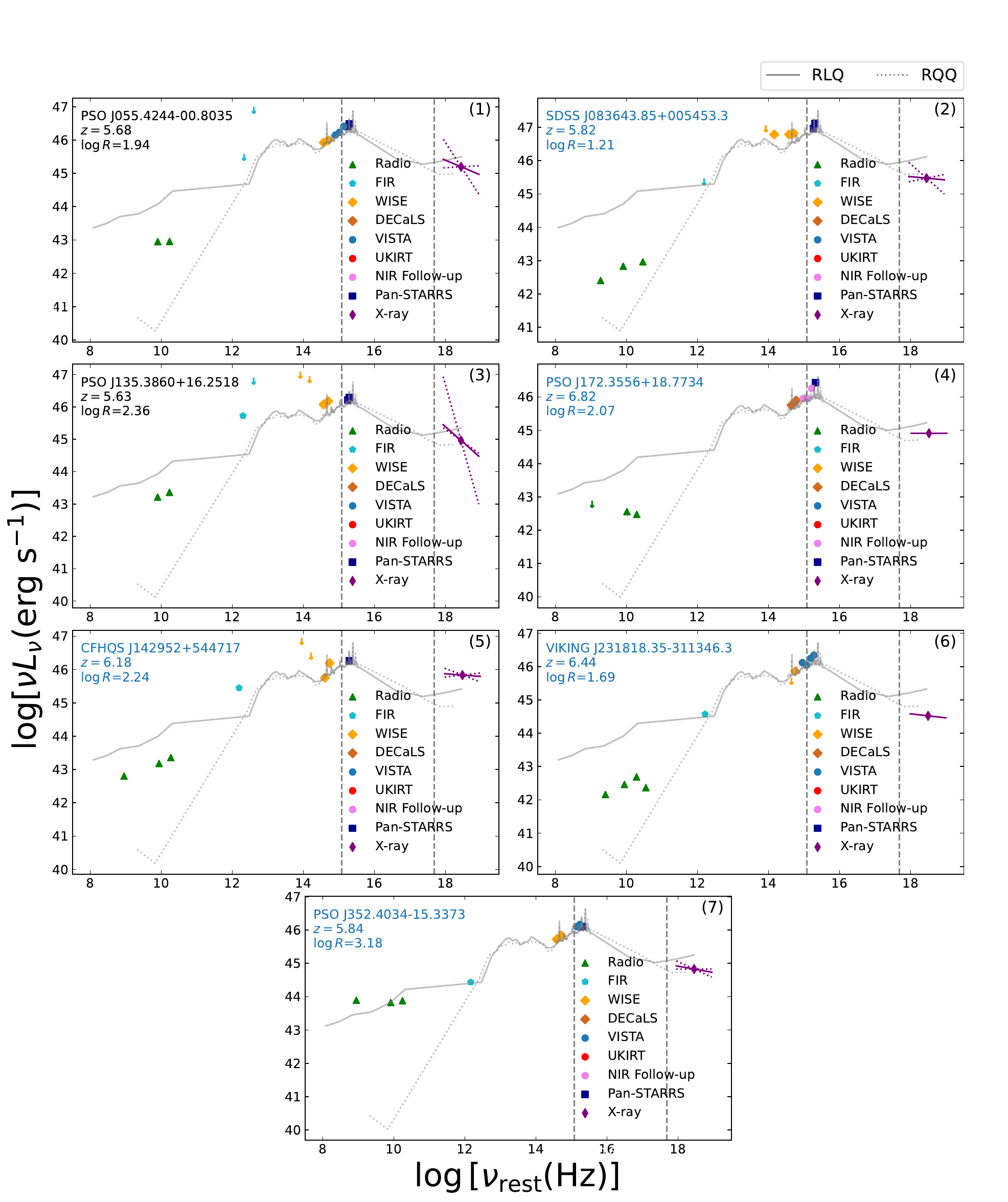}}
\caption{SEDs of the $z>5.5$ RLQ sample, ordered by ascending RA. Downward arrows represent upper limits. Objects with black text are \textit{Chandra} Cycle 23 objects. Those with blue text are archival objects. The purple diamonds are located at the observed-frame 2 keV. The solid purple lines are X-ray power-law spectra around observed-frame 2 keV derived from photon indices reported in Table 4. Their uncertainties are represented by the dotted purple lines. The solid grey curves are composite SEDs for RLQs from \citet{Shang2011}. The dotted grey lines are composite SEDs for RQQs from \citet{Shang2011}. All solid grey curves are scaled to the flux density at rest-frame 2500 \AA. The dotted grey vertical lines are at rest-frame 2500 {\AA} and 2 keV. The name, redshift, and $\log R$ of each object are shown at the upper left of each panel.   }
\label{fig:sedhighz}
\end{figure*}

 We calculated the monochromatic luminosities using flux densities and distances, which are derived from redshifts. The constructed SEDs for $4<z<5.5$ HRLQs and $z>5.5$ RLQs are shown in Fig. \ref{fig:sedlowz} and Fig. \ref{fig:sedhighz}. Both figures are ordered by ascending RA. Following \citetalias{wu13} and \citetalias{zhu19}, the composite SEDs for the 10 HRLQs from \citet{Shang2011} with comparable optical luminosity $(\log\lambda L_\lambda(3000\angstrom)>45.9)$ and radio loudness $(2.9<\log R<3.7)$ are plotted for comparison in Fig. \ref{fig:sedlowz} and Fig. \ref{fig:sedhighz}. Considering the radio-loudness of $z>5.5$ RLQs ($1.21<\log R<3.18$), we also plotted the composite spectrum for RQQs retrieved from \citet{Shang2011} using dotted lines in Fig. \ref{fig:sedhighz}. The composite SEDs are normalized to the observed fluxes at \mbox{2500 \angstrom}.
\par Four out of nine newly-added HRLQs have higher luminosities at rest-frame 2 keV (the rightmost dotted vertical grey line of each panel) than the comparison SEDs. The \mbox{X-ray} luminosities of the remaining five sources are comparable to the comparison SED if the uncertainties of $\Gamma_\mathrm{X}$ are considered. Combining the SED comparison results from \citetalias{wu13}, \citetalias{zhu19}, and our work, 20 out of 41 HRLQs show apparent \mbox{X-ray} excesses. Indeed, all 20 objects have $\Delta\alpha_\mathrm{ox,RLQ}\geq 0.05$, indicating that the SED comparison method is consistent with the method described in Section \ref{sec: X-ray, optical/UV, and radio properties}. Nearly all of them have \mbox{X-ray} luminosities that are at least comparable to the comparison SED. 
\par The rest-frame 2 keV luminosities for three out of seven $z>5.5$ RLQs are higher than the comparison composite SED of RLQs. The three objects that show apparent \mbox{X-ray} enhancements under the SED comparison are those with the largest $\Delta\alpha_\mathrm{ox,RLQ}$ values among the $z>5.5$ RLQs.   
\section{Discussion}

\subsection{Implications for the fractional IC/CMB model} 
\par Historically, it has been thought that the small-scale \mbox{X-ray} core emission of RLQs is dominated by a jet-linked component \citep[e.g.][]{Worrall1987}{}{}. To accommodate the mild \mbox{X-ray} enhancements of high-$z$ HRLQs, \citetalias{wu13} proposed a \textit{fractional} IC/CMB model, where the IC/CMB process mainly operates within a few kpc from the central SMBH and the contribution from the IC/CMB process is only a small fraction of the total \mbox{X-ray} emission at low redshifts. None of the objects in our sample shows extended X-ray emission. Considering that the 0.5 arcsec angular resolution of \textit{Chandra} corresponds to a physical size of 3.4 kpc at $z=4.35$ (the median redshift of the $4<z<5.5$ HRLQ sample), the X-ray jets (if present) are likely smaller than \mbox{$\approx5$ kpc}. This is consistent with the assumptions of the fractional IC/CMB model. \mbox{X-ray} observations with higher spatial resolution are required to constrain the scale of the jet. An alternative approach is to observe gravitationally lensed RLQs at high redshifts, whose \mbox{X-ray} emission is expected to have a significant contribution from the jet. However, only a few gravitationally lensed RLQs, limited to $z\lesssim3.6$, have been observed in X-rays. Those with the highest redshifts are either only moderately radio loud \citep[][]{Dadina2016,Dogruel2020}{}{} or have ambiguities in physical interpretation \citep[][]{Schwartz2021}{}{}. Besides, none of the gravitationally lensed systems mentioned above shows apparent X-ray enhancements compared to typical RLQs.

\par The results from \citet{zhu20,zhu21} may suggest a simpler explanation. \citet{zhu20} found that the contribution of the jet-linked component to the \mbox{X-ray} core emission of RLQs is generally small, with a mean fraction of $\approx5\%$. The median redshift of the RLQ sample utilized by \citet{zhu20} is 1.5. The small contribution from the jet-linked component could readily explain the \mbox{X-ray} enhancements of high-$z$ HRLQs. 
\par We then modeled the factor of \mbox{X-ray} enhancement as \mbox{$1+A\left[(1+z)^4-1\right]$}, containing a constant component and an IC/CMB component that evolves with redshift following $(1+z)^4$, to constrain the contribution from the IC/CMB process at different redshifts. $A$ is set to $9.41\times10^{-4}$ to obtain the \mbox{X-ray} enhancement factor of 1.77 at $z=4.35$. At $z\approx1.5$, the predicted factor of \mbox{X-ray} enhancement is $\approx4\%$, which is consistent with previous studies (e.g. \citetalias{wu13}; \citetalias{zhu19}; \citealp{Ighina2019}) and the contribution from the jet-linked component found by \citet{zhu20}. This is also roughly consistent with the jet-to-core X-ray ratio reported by \citet[][]{Marshall2005,Marshall2011,Marshall2018}{}{}.

\subsection{The \mbox{X-ray} enhancements of RLQs at $z>5.5$}
\label{sec:twopop}
At $z>5.5$, the \mbox{X-ray} emission of moderately radio-loud quasars may be enhanced as a manifestation of the IC/CMB process in the dense CMB photon field. Indeed, the $z>5.5$ RLQs show \mbox{X-ray} enhancements as expected if the IC/CMB mechanism plays an important role. However, the $z>5.5$ RLQs show properties different from the HRLQs at $4<z<5.5$.
\par We compiled multiwavelength radio data for 41 HRLQs at \mbox{$4<z<5.5$} and seven RLQs at $z>5.5$ in Table \ref{tab:radioprop}. We calculated the radio spectral index between 1.4 GHz and a lower frequency, $\alpha_\mathrm{low}$, and the radio spectral index between \mbox{1.4 GHz} and a higher frequency, $\alpha_\mathrm{high}$. We then took the average of $\alpha_\mathrm{low}$ and $\alpha_\mathrm{high}$ to get $\alpha_\mathrm{r}$, the radio spectral index around observed-frame 1.4 GHz defined as $f\propto\nu^{\alpha_\mathrm{r}}$. When calculating $\alpha_\mathrm{low}$ and $\alpha_\mathrm{high}$, we preferred to use frequencies that are further from 1.4 GHz to reduce the effects of measurement uncertainties with the exception that 150 MHz has the lowest priority since it is too far away from 1.4 GHz and the radio spectra of some objects are peaked at frequencies between 150 MHz and 1.4 GHz. The radio data we collected cover rest-frame frequencies of \mbox{$\approx0.9$--27 GHz} and $\approx1$--34 GHz for typical $4<z<5.5$ HRLQs and $z>5.5$ RLQs (calculated using the median redshift of each sample), respectively. Thus, the redshift difference between the two samples does not significantly affect the rest-frame radio frequencies probed. In Table \ref{tab:radioprop}, we used boldface to label GHz peaked sources (GPSs) identified based on the radio data we collected, whose $\alpha_\mathrm{r}$ should be treated with caution.
\par
\begin{figure*}
    \centering
    \includegraphics[scale=0.5]{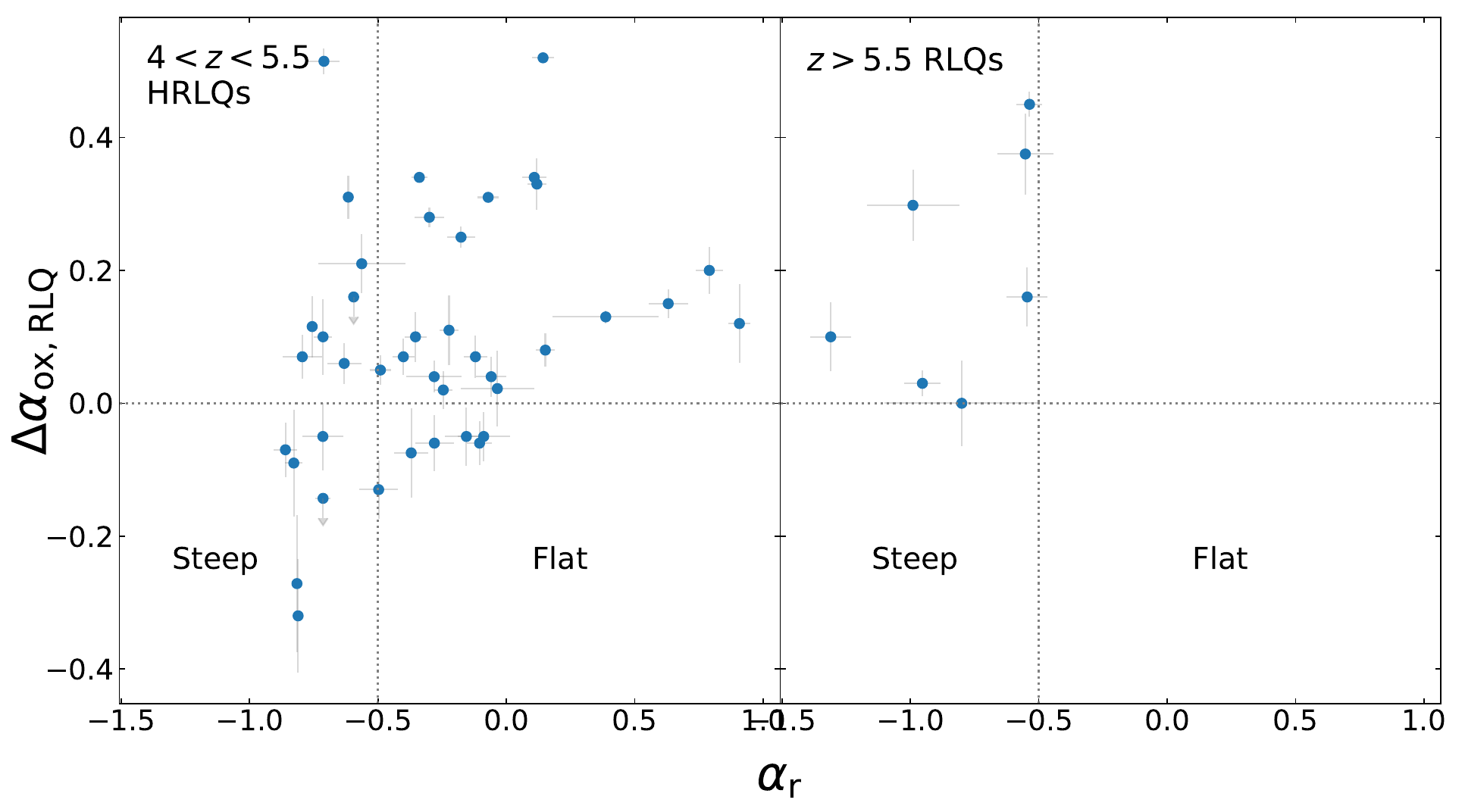}
    \caption{$\Delta\alpha_\mathrm{ox,RLQ}$ versus radio spectral index $\alpha_\mathrm{r}$ for $4<z<5.5$ HRLQs (left panel) and $z>5.5$ RLQs (right panel). The dotted horizontal line represents $\Delta\alpha_\mathrm{ox,RLQ}=0$. The dotted vertical line represents $\alpha_\mathrm{r}=-0.5$, the typical boundary between steep and flat radio spectra.}
    \label{fig:daox_alphar}
\end{figure*}

 To explore the correlation between \mbox{X-ray} enhancements and radio spectral index, we plotted $\Delta\alpha_\mathrm{ox,RLQ}$ versus $\alpha_\mathrm{r}$ in Fig. \ref{fig:daox_alphar} for \mbox{$4<z<5.5$} HRLQs (left panel) and $z>5.5$ RLQs (right panel). The medians of $\Delta\alpha_\mathrm{ox,RLQ}$ of $4<z<5.5$ HRLQs and $z>5.5$ RLQs with $\alpha_\mathrm{r}<-0.5$ are $0.06^{+0.04}_{-0.13}$ and $0.16^{+0.14}_{-0.13}$, respectively.

Our results suggest that RLQs at $z>5.5$ with $\alpha_\mathrm{r}<-0.5$ tend to have stronger \mbox{X-ray} enhancements than HRLQs at \mbox{$4<z<5.5$} with similar radio spectral indices. Considering that RLQs at \mbox{$4<z<5.5$} do not show \mbox{X-ray} enhancements \citep[e.g.][]{miller2011}{}{}, the \mbox{X-ray} enhancements of RLQs apparently rise significantly at $z\gtrsim5.5$. It is worth noting that although radio-selected blazars that should have the strongest jet-linked X-ray components have been deliberately removed (see Section \ref{sec:RLQ sample}), the X-ray enhancement of our optically selected $z>5.5$ RLQ sample is still strong.
We also note that there is no apparent correlation between $\log R$ and \mbox{X-ray} enhancements for $z>5.5$ RLQs. The only HRLQ in our $z>5.5$ RLQ sample (PSO J352$-$15) only shows a marginal \mbox{X-ray} enhancement with $\Delta\alpha_\mathrm{ox,RLQ}=0.03^{+0.02}_{-0.02}$. This does not support the idea that \mbox{X-ray} emission for $z>5.5$ RLQs is dominated by jets. However, the lack of correlation between $\log R$ and \mbox{X-ray} enhancements is only suggestive since the $z>5.5$ RLQ sample is very small.

\par
\begin{figure*}
    \centering
    \includegraphics[scale=0.5]{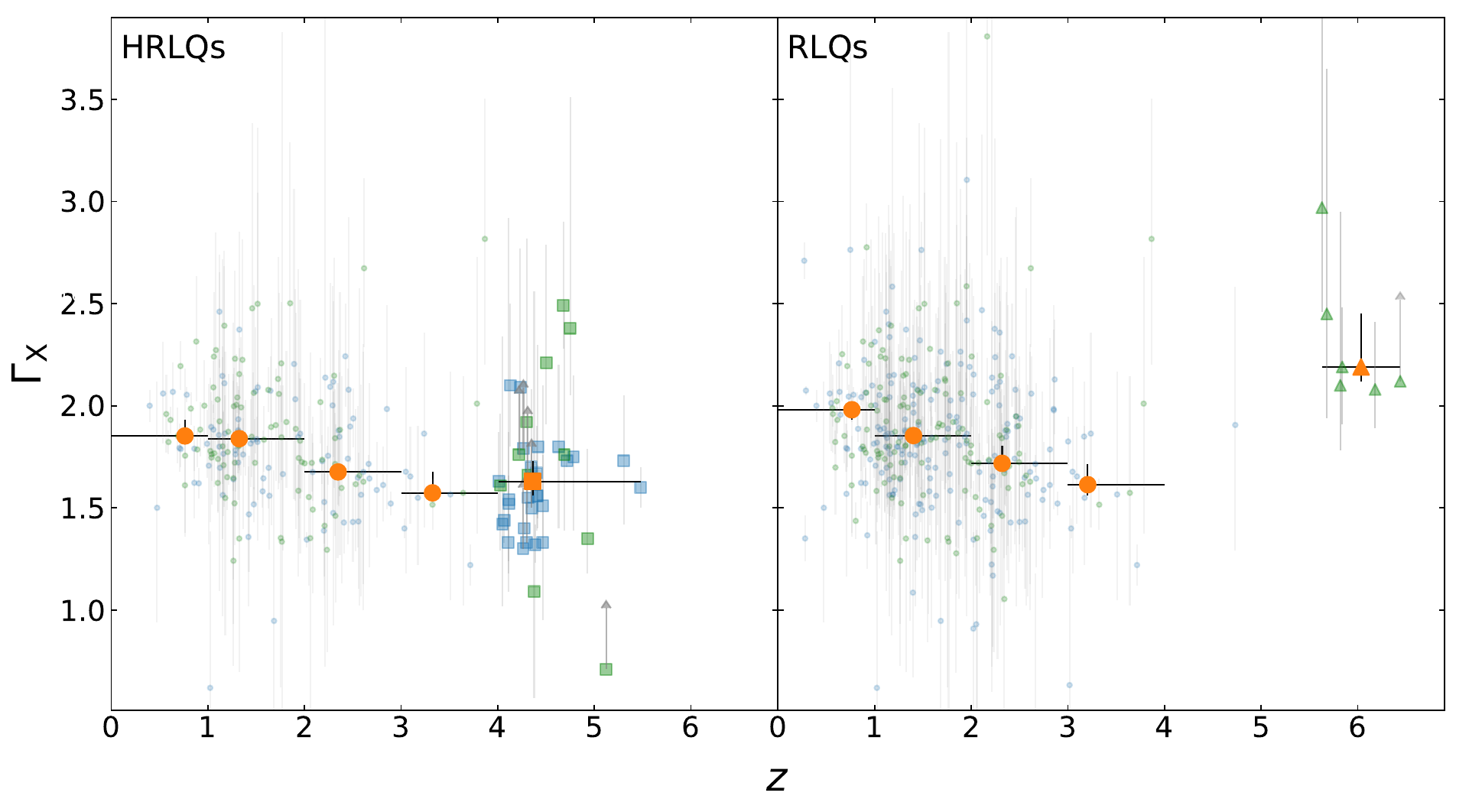}
    \caption{\mbox{X-ray} power-law photon index $\Gamma_\mathrm{X}$ versus redshift for HRLQs (left panel) and RLQs (right panel). 38 out of 41 HRLQs at $4<z<5.5$ and six out of seven RLQs at $z>5.5$ have available $\Gamma_\mathrm{X}$ values. In both panels, blue and green symbols represent objects with $\alpha_\mathrm{r}>-0.5$ and $\alpha_\mathrm{r}\leq-0.5$, respectively. The large orange dots are the median $\Gamma_\mathrm{X}$ values of objects from four redshift bins ($z=$ 0--1, 1--2, 2--3, and 3--4) in both panels; see Footnote \ref{footnote:gamma trend} for further relevant discussion. The $z$ coordinates of orange dots represent the median $z$ of redshift bins with horizontal error bars representing the ranges of redshift bins. The vertical error bars of orange symbols are 1$\sigma$ uncertainties obtained via bootstrapping. \textbf{Left:} The squares and dots (both blue and green) represent HRLQs at $4<z<5.5$ and HRLQs from \citet{zhu21}, respectively. The orange square represents the median $\Gamma_\mathrm{X}$ value of $4<z<5.5$ HRLQs. The $z$ coordinate of the orange square is the median redshift of the $4<z<5.5$ HRLQ sample with its horizontal error bar representing the redshift range of the $4<z<5.5$ HRLQs sample. \textbf{Right:} The triangles and dots (both blue and green) represent RLQs at $z>5.5$ and RLQs from \citet{zhu21}, respectively. The orange triangle is the median $\Gamma_\mathrm{X}$ value of $z>5.5$ RLQs and is slightly displaced in the horizontal direction to avoid overlapping with green triangles. The horizontal error bar of the orange triangle indicates the redshift range of $z>5.5$ RLQs with available $\Gamma_\mathrm{X}$ values.}  
    \label{fig:gamma_z}
\end{figure*}

 We also found that the \mbox{X-ray} spectra of $z>5.5$ RLQs are steeper than those of their low-redshift counterparts while such an \mbox{X-ray} steepening is not found in $4<z<5.5$ HRLQs. We plotted $\Gamma_\mathrm{X}$ versus redshift for 333 optically selected RLQs with $z\leq4.73$ analyzed by \citet{zhu21} (small dots) and median $\Gamma_\mathrm{X}$ in four redshift bins ($z=$ 0--1, 1--2, 2--3, and 3--4; large orange dots) in the right panel of Fig. \ref{fig:gamma_z}.\footnote{\label{footnote:gamma trend}
We notice that there is an apparent $\Gamma_\mathrm{X}$ versus redshift trend for $z<4$ RLQs and HRLQs as is shown in both panels of Fig. \ref{fig:gamma_z} which is probably caused by selection biases. The radio-loud nature of objects in \citet{zhu21} is determined using radio data at observed-frame 1.4 GHz, whose rest-frame frequency increases with redshift. At higher rest-frame frequencies,  objects with steep radio spectra may drop below detection limits. Thus, objects in the \citet{zhu21} sample tend to have flatter radio spectra at higher redshifts, which suggests more jet contribution and hence flatter \mbox{X-ray} spectra. Aside from that, only bright radio sources can be detected at higher redshifts, resulting in higher $\log R$ values and probably flattening \mbox{X-ray} spectra.
} The \citet{zhu21} sample was built upon the \citet{zhu20} sample by selecting objects with high-quality X-ray data that could produce reliable $\Gamma_\mathrm{X}$ values. Throughout Fig. \ref{fig:gamma_z}, blue labels represent objects with $\alpha_\mathrm{r}>-0.5$ and green labels represent those with $\alpha_\mathrm{r}\leq -0.5$. The median$(\sigma_\mathrm{NMAD})$ of $\Gamma_\mathrm{X}$ for low-redshift RLQs given by \citet{zhu21} is $1.84^{+0.01}_{-0.01}(0.30)$. We then included the six $z>5.5$ RLQs with available $\Gamma_\mathrm{X}$ values (small triangles) and their median $\Gamma_\mathrm{X}$ (large orange triangle) in Fig. \ref{fig:gamma_z} for comparison. The potentially heavily absorbed object, PSO J172+18 (see Section \ref{sec: XMM reduction}), was not included. We used the Kaplan-Meier estimator to derive the median$(\sigma_\mathrm{NMAD})$ of $\Gamma_\mathrm{X}$ for $z>5.5$ RLQs, $2.19_{-0.07}^{+0.26}(0.10)$, which is significantly larger than that for low redshift RLQs.
\par The small dots in the left panel of Fig. \ref{fig:gamma_z} are HRLQs from the \citet{zhu21} sample and the large orange dots are median $\Gamma_\mathrm{X}$ values of HRLQs in redshift bins similar to the right panel. The squares represent $4<z<5.5$ HRLQs. The large yellow square represents the median $\Gamma_\mathrm{X}$ value of $4<z<5.5$ HRLQs. 38 out of 41 objects in our $4<z<5.5$ HRLQ sample have available $\Gamma_\mathrm{X}$ values. The median$(\sigma_\mathrm{NMAD})$ of $\Gamma_\mathrm{X}$ for HRLQs at $z<4$ and $4<z<5.5$ are $1.80^{+0.08}_{-0.09}(0.26)$ and $1.63_{-0.07}^{+0.10}(0.27)$, respectively. The 1$\sigma$ error bars were calculated using bootstrap. The \mbox{X-ray} spectra of HRLQs at $4<z<5.5$ are not steeper compared to their low-redshift counterparts. The steepening of \mbox{X-ray} spectra of $z>5.5$ RLQs and the lack of it for $4<z<5.5$ HRLQs indicate $z>5.5$ RLQs may be in a different evolutionary stage.
\par The evolution of \mbox{X-ray} spectral properties of RQQs up to $z\gtrsim6$ has been investigated by recent studies. \citet{Vito2019} jointly fitted the \mbox{X-ray} spectra of 12 RQQs at $z>6$ with $<30$ net counts and 6 with \mbox{$>30$} net counts from 0.5--7 keV. They found an average $\Gamma_\mathrm{X}=2.20^{+0.39}_{-0.34}$ for the sample with $<30$ net counts and an average $\Gamma_\mathrm{X}=2.13^{+0.08}_{-0.08}$ for the sample with \mbox{$>30$} net counts. \citet{Zappacosta2023} reported an average $\Gamma_\mathrm{X}=2.4^{+0.1}_{-0.1}$ for a sample of $z>6$ hyperluminous RQQs ($L_\mathrm{bol}\geq10^{47}$ erg s$^{-1}$). Compared to the canonical value of $\Gamma_\mathrm{X}\approx1.9$ for RQQs at lower redshifts, both studies suggest, although at different significance levels, the steepening of \mbox{X-ray} spectra of $z\gtrsim6$ RQQs. \citet{Zappacosta2023} proposed that the steepening of \mbox{X-ray} spectra is a manifestation of lower coronal temperature due to disc truncation caused by radiatively driven winds. The $\Gamma_\mathrm{X}$ evolution trend of RQQs (see also Fig. 7 in \citealp{Vito2019} and Fig. 5 in \citealp{Zappacosta2023}) is similar to that of RLQs presented in the right panel of \mbox{Fig. \ref{fig:gamma_z}}. This suggests that RLQs may undergo similar physical transitions as RQQs at $z\sim 6$.

\par The \mbox{X-ray} spectral and other properties discussed above indicate $z>5.5$ RLQs are at a different evolutionary stage. RLQs at $z>5.5$ feature steep radio and \mbox{X-ray} spectra and relatively strong \mbox{X-ray} enhancements. The \mbox{X-ray} steepening mechanism proposed by previous studies, which was mentioned in the previous paragraph, could probably shed light on the evolution of RLQs. However, a larger sample of $z>5.5$ RLQs is needed to determine the physical mechanisms that are responsible for the \mbox{X-ray} enhancements of RLQs at very high redshifts and the role of the IC/CMB process in them.

\subsection{Milli-arcsecond scale properties}
We also investigated if there is a relation between the milliarcsec-scale radio-jet properties and the \mbox{X-ray} enhancements of high redshift HRLQs/RLQs. We used the ratio between the flux densities given by VLBI observations and observations with lower resolution (NVSS/FIRST) at matched frequencies, $f_\mathrm{VLBI}/f_\mathrm{lowres}$, as an indicator of source compactness. An $f_\mathrm{VLBI}/f_\mathrm{lowres}$ that is close to one indicates a compact radio-emitting region while a ratio that is much smaller than one indicates the existence of diffuse emission extending beyond the scale of a few hundred pc.
\par We collected available VLBI observations from the literature (see Table \ref{tab:radioprop}). 29 out of 41 HRLQs at $4<z<5.5$ and five out of seven RLQs at $z>5.5$ have available VLBI observations. If an object has VLBI observations in more than one band, the band closest to observed-frame 1.4 GHz was selected. The frequencies of all the bands we selected are greater than 1.4 GHz. We then used $\alpha_\mathrm{high}$ (see Table \ref{tab:radioprop}) to extrapolate the NVSS/FIRST flux densities from 1.4 GHz to the VLBI frequencies we selected. Since the VLBI and NVSS/FIRST observations were not taken simultaneously, variability of the sources may influence the values of $f_\mathrm{VLBI}/f_\mathrm{lowres}$. Based on the VLASS radio variability analyses presented in Section \ref{sec:sample selection}, the typical radio variability of $4<z<5.5$ HRLQs and $z>5.5$ RLQs is within $\approx30$\% over a rest-frame period of \mbox{$\approx5$--6 months}. Although some VLBI and NVSS/FIRST observations are separated by over a decade in the observed-frame ($\gtrsim2$ years in the \mbox{rest-frame}), the available VLASS radio variability analyses over a shorter time period do not suggest strong variability effects upon $f_\mathrm{VLBI}/f_\mathrm{lowres}$.
\par
\begin{figure}
\centering
\includegraphics[width=\columnwidth]{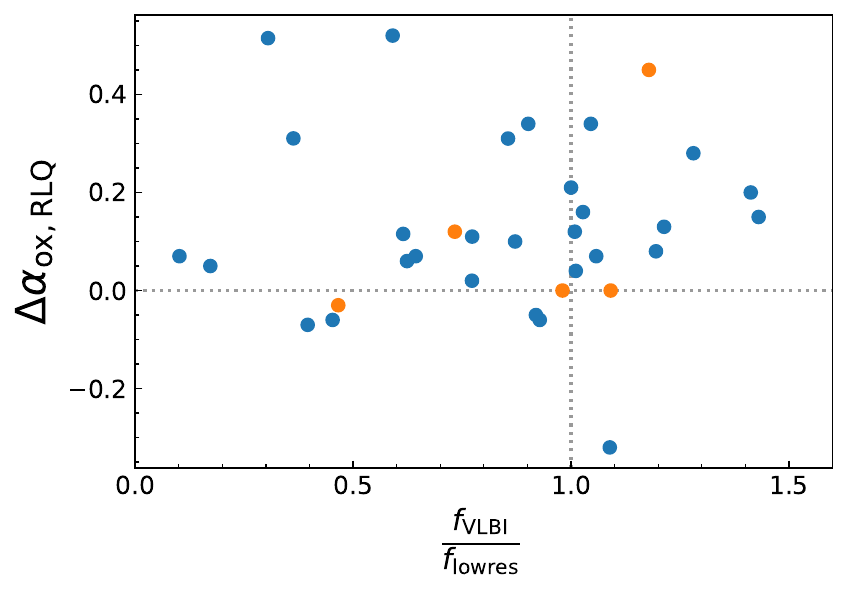}
\caption{X-ray enhancement strength versus compactness of the $4<z<5.5$ HRLQ sample and the $z>5.5$ RLQ sample. 29 out of 41 $4<z<5.5$ HRLQs (blue dots) and five out of seven $z>5.5$ RLQs (orange dots) have VLBI coverage. $f_\mathrm{VLBI}$ is the observed flux density at the VLBI band closest to 1.4 GHz. $f_\mathrm{lowres}$ is the flux density at corresponding VLBI bands inferred from $f_\mathrm{1.4GHz}$ assuming the radio spectral slope to be $\alpha_\mathrm{high}$. The vertical and horizontal dotted lines represent $f_\mathrm{VLBI}/f_\mathrm{lowres}=1$ and $\Delta\alpha_\mathrm{ox,RLQ}=0$, respectively. There is no apparent correlation between X-ray enhancement and compactness.}
\label{fig:compactness}
\end{figure}

 We plotted $\Delta\alpha_\mathrm{ox}$ versus $f_\mathrm{VLBI}/f_\mathrm{lowres}$ for $4<z<5.5$ HRLQs and $z>5.5$ RLQs that have VLBI observations in Fig. \ref{fig:compactness}. We did not find an apparent dependence between the amount of \mbox{X-ray} enhancement and the source compactness, or the scale and size of radio jets.
\par Since the $f_\mathrm{VLBI}/f_\mathrm{lowres}$ value reaches as high as $\approx1.5$, we expect that source variability and flux uncertainties may affect $f_\mathrm{VLBI}/f_\mathrm{lowres}$ by $\approx0.5$. This is roughly consistent with a combination of the suggestive $\lesssim30\%$ radio variability mentioned in the previous paragraph and measurement uncertainties, considering that objects with longer time gaps between VLBI and NVSS/FIRST observations may show stronger radio variability. It is thus appropriate to consider objects with $f_\mathrm{VLBI}/f_\mathrm{lowres}>0.5$ as relatively compact. The majority of objects in Fig. \ref{fig:compactness} are considered to be compact and have radio jets at scales smaller than a few hundred pc, which are smaller than typical \mbox{X-ray} jets. This is consistent with the expectation that \mbox{X-ray} jets produced via the IC/CMB process are larger than radio jets since the electrons that are responsible for \mbox{X-ray} emission have a longer cooling time than those responsible for radio emission \citep[e.g.][]{Harris2007,worrall2009}{}{}.

\section{summary and future work}

\subsection{Summary}
We have investigated and confirmed the \mbox{X-ray} enhancements of $z>4$ HRLQs compared to their low-redshift counterparts using new and archival \textit{Chandra} observations. We also found such \mbox{X-ray} enhancements in $z>5.5$ RLQs for the first time. Our key findings are summarized as the following:
\begin{enumerate}
    \item We combined the $4<z<5.5$ HRLQs selected by \citetalias{wu13} and \citetalias{zhu19} to construct a sample of 41 HRLQs at $4<z<5.5$ with complete \mbox{X-ray} coverage, pushing the optical flux limit to $m_i=21.33$ compared with previous studies (See Section \ref{sec:HRLQ sample}). We obtained new \textit{Chandra} Cycle 23 observations of six HRLQs at $4<z<5.5$. Along with three archival objects, we analyzed the \textit{Chandra} observations of nine HRLQs at $4<z<5.5$ and reported their \mbox{X-ray} properties in Table \ref{tab:lowzresults} (see Section \ref{sec:chandra analyses}). 
    \item We searched the literature and constructed a sample of optically selected $z>5.5$ RLQs. The sample consists of the seven brightest (in rest-frame optical/UV bands) optically selected RLQs at $z>5.5$ with complete X-ray coverage (see Section \ref{sec:RLQ sample}).  We also obtained new \textit{Chandra} Cycle 23 observations of two RLQs at $z>5.5$. Along with five archival objects, we analyzed and reported the \mbox{X-ray} properties of seven RLQs at $z>5.5$ in Table \ref{tab:highzresults}. Among the five archival objects, two (VIK J2318$-$3113 and PSO J172+18) have not been reported in the literature previously. The remaining three objects have been reanalyzed for consistency. All objects except for PSO J172+18, which only has \textit{XMM-Newton} coverage (see Section \ref{sec: XMM reduction}), were observed by \textit{Chandra} (see Section \ref{sec:chandra analyses}).
    \item The \mbox{X-ray} enhancement of HRLQs at $4<z<5.5$ compared to matched HRLQs at $z<4$ has been further confirmed. The $\Delta\alpha_\mathrm{ox}$ distributions of $4<z<5.5$ HRLQs are shown to be different from those of $z<4$ HRLQs at a 4.90--5.33$\sigma$ statistical significance level. The factor of \mbox{X-ray} enhancement of $4<z<5.5$ HRLQs is constrained to be $1.77^{+0.19}_{-0.12}$ (see Section \ref{sec:results-HRLQ}).
    \item We found an \mbox{X-ray} enhancement of RLQs at $z>5.5$ compared to matched RLQs at $z\lesssim5$ for the first time. The $\Delta\alpha_\mathrm{ox}$ distributions of $z>5.5$ RLQs are shown to be different from those of $z\lesssim5$ RLQs at a 3.65--4.91 $\sigma$ statistical significance level. The factor of \mbox{X-ray} enhancement of $z>5.5$ RLQs is $2.72^{+3.58}_{-1.47}$ (see Section \ref{sec:results-RLQ}).
    \item We constructed broadband SEDs of the nine newly added $4<z<5.5$ HRLQs and seven $z>5.5$ RLQs. We compared these with suitable composite SEDs at lower redshifts. The comparison results further support the \mbox{X-ray} enhancements of $4<z<5.5$ HRLQs and $z>5.5$ RLQs (see Section \ref{sec:sed}). The broadband SEDs of the remaining 32 HRLQs at $4<z<5.5$ can be found in \citetalias{wu13} and \citetalias{zhu19}.
    \item We found that the RLQs at $z>5.5$ have a median X-ray power-law spectral index ($\Gamma_\mathrm{X}$) of $2.19^{+0.26}_{-0.07}$, which is significantly steeper than that of RLQs at lower redshifts, $1.84^{+0.01}_{-0.01}$, given by \citet{zhu21}.
\end{enumerate}

\subsection{Future work}
\par The number of luminous RLQs at very high redshifts ($z\gtrsim6$) has been increasing rapidly with the aid of wide-field sky surveys in the optical-to-NIR bands (e.g. Pan-STARRS, \citealp{Chambers2016}; UKIDSS, \citealp{UKIDSS}; and VHS, \citealp{VHS}) and at radio frequencies (e.g. RACS, \citealp{McConnell2020}; and LoTSS-DR2, \citealp{shimwell2022}). During the writing process of this paper, several new RLQs at $z\gtrsim6$ have been identified \citep[][]{Banados2023,Ighina2023}{}{}. The upcoming Vera C. Rubin Observatory Legacy Survey of Space and Time \citep[][]{Ivezic2019}{}{}, together with \textit{Euclid} and the \textit{Roman Space Telescope}, are expected to discover even more quasars, pushing the redshift frontier of quasars to \mbox{$z\approx9$--10}. Their radio properties could be constrained by ongoing sensitive radio surveys including VLASS \citep{lacy2020} and EMU \citep{Norris2021}. Thus, the number of RLQs at $z\gtrsim6$ is expected to be significantly increased in the near future. With the increasing number of RLQs, one could possibly form a sample of HRLQs at $z\gtrsim6$ as well.
\par While our results for $4<z<5.5$ HRLQs can still be explained by the fractional IC/CMB model, it is possible that there are additional mechanisms that are responsible for the \mbox{X-ray} enhancement of \mbox{$z>5.5$} RLQs. Our understanding of RLQs/HRLQs at $z\gtrsim6$ is greatly limited by low \mbox{X-ray} count numbers and the small sample size. Deeper \mbox{X-ray} observations of objects in our $z>5.5$ sample will be valuable to set tighter constraints on their \mbox{X-ray} spectral properties. In addition, investigating the \mbox{X-ray} properties of larger samples of $z\gtrsim6$ RLQs/HRLQs with \textit{Chandra} and \textit{XMM-Newton} could provide insights into the \mbox{X-ray} emission mechanisms of RLQs/HRLQs and the evolution of SMBHs in the first billion years of the Universe.

\section*{Acknowledgements}
We thank the second referee for constructive feedback. We acknowledge support from the Chandra X-ray Center grant GO2-23083X 
and Penn State ACIS Instrument Team Contract SV4-74018 (issued by the 
Chandra X-ray Center, which is operated by the Smithsonian Astrophysical 
Observatory for and on behalf of NASA under contract NAS8-03060). S.F.Z. and Y.Q.X. acknowledge support from NSFC grants (12025303 and 11890693). The 
Chandra ACIS Team Guaranteed Time Observations (GTO) utilized were 
selected by the ACIS Instrument Principal Investigator, Gordon P. Garmire, 
currently of the Huntingdon Institute for X-ray Astronomy, LLC, which is 
under contract to the Smithsonian Astrophysical Observatory via 
Contract SV2-82024.

\section*{Data Availability}
The raw data of all objects except SDSS J104742.57+094744.9 are publicly available in the \textit{Chandra} Data Archive (\url{https://cxc.cfa.harvard.edu/cda/}) and the \textit{XMM-Newton} Science Archive (\url{http://nxsa.esac.esa.int/nxsa-web/#home}). The raw data of SDSS J104742.57+094744.9 will be shared on reasonable request to the corresponding author. The reduced data generated in this research will be shared on reasonable request to the corresponding author.

\bibliographystyle{mnras}
\bibliography{main}

\begin{table*}
\centering
\caption{X-ray observation log of objects analyzed in this paper}
\label{tab:obslog}
\begin{tabular}{ccccccccc}

\hline

 Object name &  RA & DEC. &  Instr. & $z$ & Obs. date & Obs. ID & Exp.time \\  &  (deg)  &  (deg) &  &  &  &   &  (ks)   \\
\hline

\multicolumn{8}{c}{\textit{Chandra} Cycle 23 targets}\\
PSO J055.4244$-$00.8035  & 55.4244 & $-0.8035$ & ACIS-S  & 5.68  & 2022/02/03 & 26029 & 13.18\\
SDSS J082511.60+123417.2 &  126.2983 & 12.5714 & ACIS-S  & 4.377 & 2021/10/17 & 26025 & 6.20\\
PSO J135.3860+16.2518    &  135.3860 & 16.2519 & ACIS-S  & 5.63  & 2022/01/26 & 26030 & 12.74\\
SDSS J104742.57+094744.9 &  161.9274 &  9.7958 & ACIS-S  & 4.233 & 2023/03/12 & 26023 & 5.02\\
SDSS J115605.44+444356.5 &  179.0223 & 44.7301 & ACIS-S  & 4.310 & 2021/12/21 & 26027 & 7.78\\
SDSS J125300.15+524803.3 &  193.2506 & 52.8009 & ACIS-S  & 4.115 & 2021/11/10 & 26028 & 9.17\\
SDSS J153830.71+424405.6 &  234.6280 & 42.7349 & ACIS-S  & 4.099 & 2021/12/05 & 26026 & 6.67\\
SDSS J165539.74+283406.7 &  253.9156 & 28.5685 & ACIS-S  & 4.048 & 2021/12/05 & 26024 & 5.10\\

\multicolumn{8}{c}{Archival objects}\\
SDSS J083643.85+005453.3 & 129.1829 & 0.9148  & ACIS-S  & 5.82  & 2002/01/29 & 3359  & 5.68 \\
SDSS J094004.80+052630.9 & 145.0200 & 5.4419  &  ACIS-S & 4.503 & 2017/12/31 & 20476 & 5.99\\
PSO J172.3556+18.7734    & 172.3556 & 18.7734 & EPIC-pn & 6.82  & 2020/12/20 & 0863780101 & 94.03\\
                         &          &         &         &       & 2020/12/22 & 0863780201 & 89.73\\
SDSS J140025.40+314910.6 & 210.1058 & 31.8196 & ACIS-S  & 4.690 & 2018/02/07 & 20480 & 5.99\\
CFHQS J142952+544717     & 217.4671 & 54.7882 & ACIS-S  & 6.18  & 2021/08/03 & 22601 & 30.56\\
SDSS J154824.01+333500.1 & 237.1000 & 33.5834 & ACIS-S  & 4.678 & 2018/02/11 & 20482 & 6.11 \\
VIKING J231818.35$-$311346.3 & 349.5765 &$-$31.2295 &ACIS-S &6.44 & 2022/01/07 & 25739 & 19.64 \\ 
& & & & & 2022/01/08 & 26254 & 9.78 \\
& & & & & 2022/04/12 & 25254 & 9.94\\
& & & & & 2022/04/17 & 26391 & 17.84 \\
& & & & & 2022/04/18 & 26392 & 9.94 \\
PSO J352.4034$-$15.3373  & 352.4034 & -15.3373 & ACIS-S & 5.83 & 2019/08/19 & 21415 & 41.52 \\
& & & & & 2019/08/21 & 22728 & 59.28 \\
& & & & & 2019/08/24 & 22729 & 45.46 \\
& & & & & 2019/08/25 & 22730 & 38.24 \\
& & & & & 2019/09/16 & 21416 & 19.06 \\
& & & & & 2019/09/17 & 22850 & 31.44 \\
& & & & & 2019/09/22 & 22851 & 29.88 \\
\hline 

\end{tabular}
\end{table*}

\begin{table*}
\centering
\caption{X-ray net counts, hardness ratios, and effective photon indices of objects analyzed in this paper}
\label{tab:cnts}
\begin{threeparttable}[b]
\begin{tabular}{cccccc}
\hline
 Name & \multicolumn{3}{c}{Net X-ray counts} & HR$^\mathrm{a}$ & $\Gamma_\mathrm{X}$ \\
\hline
{}& Full band   &  Soft band   &  Hard band  &{} &{}\\
{}&   (0.5--8 keV)  &    (0.5--2 keV) &    (2--8 keV)  &{} &{}\\

\multicolumn{6}{c}{\textit{Chandra} Cycle 23 targets}\\
PSO J055.4244$-$00.8035 & $12.6^{+3.8}_{-3.2}$ & $8.5^{+3.2}_{-2.6}$ & $4.1^{+2.4}_{-1.7}$ & $0.49^{+0.26}_{-0.31}$ & $2.45^{+1.20}_{-0.51}$ \\
SDSS J082511.60+123417.2 &  $5.2^{+2.6}_{-2.0}$ & $2.1^{+1.8}_{-1.2}$ & $3.1^{+2.1}_{-1.5}$  &    $1.49_{-1.31}^{+0.88}$& $1.09^{+1.47}_{-0.52}$ \\
PSO J135.3860+16.2518 & $8.3^{+3.2}_{-2.6}$ &  $6.3^{+2.8}_{-2.2}$ & $2.0^{+1.8}_{-1.2}$ & $0.31^{+0.16}_{-0.30}$ &  $2.97_{-0.51}^{+3.77}$ \\
SDSS J104742.57+094744.9 & $27.7^{+5.4}_{-4.8}$ & $16.0^{+4.2}_{-3.6}$ & $11.7^{+3.7}_{-3.0}$ & $0.73^{+0.24}_{-0.31}$ & $2.09^{+0.68}_{-0.35}$\\
SDSS J115605.44+444356.5 &      ${5.1}^{+2.8}_{-2.1}$ &$4.2^{+2.6}_{-1.8}$ &  $<4.0$  &   $<0.94$ & $>1.66$\\
SDSS J125300.15+524803.3 &      ${10.5}^{+3.5}_{-2.9}$ & $5.2^{+2.6}_{-2.0}$ & $5.2^{+2.6}_{-2.0}$  &  $1.00_{-0.69}^{+0.45}$& $1.52^{+1.40}_{-0.43}$ \\
SDSS J153830.71+424405.6 &     $<2.46$ & $<2.46$ &   $<2.46$  & - &       - \\
SDSS J165539.74+283406.7 &  $15.8^{+4.2}_{-3.6}$ & $ 7.4^{+3.0}_{-2.4} $& $8.4^{+3.2}_{-2.6}$  &  $1.14_{-0.62}^{+0.47}$& $1.42^{+0.92}_{-0.40}$ \\
\multicolumn{6}{c}{Archival objects}\\
SDSS J083643.85+005453.3 & $24.4^{+5.1}_{-4.5}$ & $20.2^{+4.7}_{-4.1}$ & $4.2^{+2.4}_{-1.7}$ & $0.21^{+0.08}_{-0.13}$ & $2.10^{+0.87}_{-0.31}$\\
SDSS J094004.80+052630.9 &     $37.1^{+6.7}_{-6.0}$ & $26.6^{+5.7}_{-5.0}$ &  $10.5^{+3.8}_{-3.1}$  & $0.39_{-0.16}^{+0.13}$ & $2.21^{+0.58}_{-0.30}$\\
PSO J172.3556+18.7734 $^\mathrm{b}$    &  $102.1^{+21.1}_{-21.0}$ &$38.2^{+14.0}_{-13.2}$  & $64.8^{+16.2}_{-16.1}$ &  $1.62^{+2.23}_{-0.35}$ & $1.17^{+0.23}_{-0.79}$\\
SDSS J140025.40+314910.6 &     $20.1^{+5.0}_{-4.3}$ &  $12.8^{+4.1}_{-3.4}$ &  $7.3^{+3.2}_{-2.5}$  & $0.58_{-0.29}^{+0.23}$ & $1.76^{+0.76}_{-0.37}$\\
CFHQS J142952+544717     &   $103.7_{-9.8}^{+9.9}$ & $63.7_{-7.7}^{+7.7}$ & $40.0_{-5.9}^{+6.5}$ & $0.63^{+0.11}_{-0.15}$ & $2.08^{+0.33}_{-0.19}$ \\
SDSS J154824.01+333500.1 &     $83.2^{+8.9}_{-8.8}$ & $64.0^{+7.8}_{-7.7}$ & $19.1^{+4.6}_{-4.0}$  & $0.30_{-0.10}^{+0.06}$ & $2.49^{+0.41}_{-0.21}$ \\
VIKING J231818.35$-$311346.3&  $7.9^{+3.4}_{-2.8}$ & $6.0^{+2.8}_{-2.2}$  & $<3.9$ & $<0.65$  & $>2.12$ \\
PSO J352.4034$-$15.3373  &   $122.1^{+11.0}_{-10.9}$ & $83.2^{+9.0}_{-8.9}$ & $38.8^{+6.7}_{-6.0}$ & $0.47^{+0.08}_{-0.11}$ & $2.19^{-0.19}_{+0.30}$ \\
\hline

\end{tabular}
\begin{tablenotes}
    \item[a] HR=$\frac{H}{S}$, where $H$ and $S$ are hard-band and soft-band counts, respectively. The HR values are not directly comparable across different \textit{Chandra} observation cycles (e.g., vs. \citetalias{wu13} and \citetalias{zhu19}) due to the ACIS low energy QE degradation.
    \item[b] The full, soft, and hard bands for PSO J172+18 are defined to be 0.3--4.5 keV, 0.3--1 keV, and 1--4.5 keV, respectively. See Section \ref{sec: XMM reduction}
\end{tablenotes}
\end{threeparttable}
\end{table*}

\begin{landscape}
\begin{table}
\centering
\caption{X-ray, optical/UV, and radio properties of newly added $4<z<5.5$ HRLQs. Objects are sorted by ascending RA. (1) Object name. (2) Apparent $i$-band magnitude. (3) Absolute $i$-band magnitude. (4) Galactic neutral hydrogen density \citep{HI4PI2016} in units of $10^{20}\text{ cm}^{-2}$. (5) Count rate in the observed-frame 0.5--2 keV band in units of $10^{-3}\text{ s}^{-1}$. (6) Galactic absorption-corrected flux in the observed 0.5--2 keV band, in units of $10^{-14}\text{ erg cm}^{-2}\text{ s}^{-1}$. (7) Flux density at $2/(1+z)$ keV in units of $10^{-32}$ \si{erg.cm^{-2}.s^{-1}.Hz^{-1}}. (8) Effective X-ray power-law photon index. If there is no available $\Gamma$, we take $\Gamma=1.5$ (9) Flux density observed at $2500(1+z)$ \si{\angstrom}  in units of $10^{-27}$\si{erg.cm^{-2}.s^{-1}.Hz^{-1}}. (10)  Logarithm of the monochromatic UV luminosity at rest frame 2500 \si{\angstrom} in units of \si{erg.s^{-1}.Hz^{-1}}. (11) Radio spectral index calculated near observed 1.4 \si{GHz} (see Section 5.1 for details). (12) Logarithm of monochromatic radio luminosity at rest-frame 5 \si{GHz} in units of \si{erg.s^{-1}.Hz^{-1}}, extrapolated from the observed-frame 1.4 GHz flux density assuming a power-law spectrum $f\propto\nu^{\alpha_\mathrm{r}}$. (13) Logarithm of the radio-loudness parameter. (14) Optical--X-ray spectral index. (15) The difference between the measured $\alpha_\mathrm{ox}$ and that expected for RQQs with similar UV luminosity. (16) The difference between the measured $\alpha_\mathrm{ox}$ and the expected RLQs with similar UV luminosity.}
\label{tab:lowzresults}
\begin{threeparttable}[b]
\begin{tabular}{lccccccccccccccc}
\hline
  Object name & $m_i$ & $M_i$ & $N_\mathrm{H}$ & C.R. & $F_\mathrm{X}$ & $f_{\mathrm{2keV}}$ & $\Gamma_\mathrm{X}$ &   $f_{\mathrm{2500\angstrom}}$ & $\log L_{\mathrm{2500\angstrom}}$ & $\alpha_\mathrm{r}$ & $\log L_\mathrm{r}$ & $\log R$ &   $\alpha_\mathrm{ox}$ &   $\Delta\alpha_\mathrm{ox,RQQ}$ & $\Delta\alpha_\mathrm{ox,RLQ}$ \\
  (1) & (2) & (3) & (4) & (5) & (6) & (7) & (8) & (9) & (10) & (11) & (12) & (13) & (14) & (15) & (16)\\
  \hline
\multicolumn{16}{c}{\textit{Chandra} Cycle 23 objects}\\
SDSS J082511.60+123417.2 & 20.71 & $-26.44$ & 4.47 & $0.34_{-0.20}^{+0.32}$  & $0.34^{+0.29}_{-0.19}$    & $1.03^{+0.88}_{-0.59}$    & $1.09^{+1.47}_{-0.52}$ & 0.27 & 30.97 & $-0.82$ & 33.92 & 2.82 & $-1.69^{+0.10}_{-0.14}$    &  $-0.06^{+0.10}_{-0.14}$ & $-0.27^{+0.10}_{-0.14}$      \\
SDSS J104742.57+094744.9 & 20.32 & $-26.67$ & 2.51 & $3.19^{+0.84}_{-0.71}$  & $4.91^{+1.29}_{-1.09}$    &$41.91^{+11.04}_{-9.33}$   & $2.09^{+0.68}_{-0.35}$ & 0.35 & 31.07 & $0.17$  & 33.78 & 2.59 & $-1.12^{+0.04}_{-0.04}$    &  $0.52^{+0.04}_{-0.04}$  & $0.33^{+0.04}_{-0.04}$\\
SDSS J115605.44+444356.5 & 21.06 & $-25.99$ & 1.24 & $0.56_{ -0.24}^{+0.34}$ & $0.61^{+0.38}_{-0.26}$    & $3.44^{+2.09}_{-1.47}$    & $>1.66$                & 0.18 & 30.79 & $-0.83$ & 34.50 & 3.59 & $-1.43^{+0.08}_{-0.09}$    &  $0.18^{+0.08}_{-0.09}$  & $-0.09^{+0.08}_{-0.09}$        \\
SDSS J125300.15+524803.3 & 21.33 & $-25.54$ & 1.20 & $0.57_{-0.24}^{+0.32}$  & $0.64^{+0.32}_{-0.25}$    & $3.05^{+1.53}_{-1.17}$    & $1.52^{+1.40}_{-0.43}$ & 0.13 & 30.61 & $-0.37$ & 34.31 & 3.58 & $-1.39^{+0.07}_{-0.08}$    &  $0.19^{+0.07}_{-0.08}$  & $-0.07^{+0.07}_{-0.08}$          \\
SDSS J153830.71+424405.6 & 20.76 & $-26.13$ & 2.36 & $<0.40$                 & $<0.42$ & $<1.98$ & -                      & 0.22 & 30.85 & $-0.71$ & 33.68 & 2.71 &  $<-1.56$  & $<0.05$  & $<-0.14$       \\
SDSS J165539.74+283406.7 & 20.41 & $-26.49$ & 4.58 & $1.45_{-0.47}^{+0.62}$  & $1.58^{+0.64}_{-0.51}$    & $6.75^{+2.74}_{-2.19}$    & $1.42^{+0.92}_{-0.40}$ & 0.32 & 30.99 & $-0.04$ & 33.86 & 2.75 & $-1.41^{+0.06}_{-0.07}$    &  $0.22^{+0.06}_{-0.07}$  & $0.02^{+0.06}_{-0.07}$     \\
\multicolumn{16}{c}{Archival objects}\\
SDSS J094004.80+052630.9 & 20.77& $-26.52$ & 4.07  & $4.44_{-0.83}^{+0.95}$  & $4.08 ^{+0.87}_{-0.78}$   & $41.40 ^{+8.87}_{-7.78}$  & $2.21^{+0.58}_{-0.30}$ & 0.28 & 30.98 & $-0.72$ & 34.43 & 3.32 & $-1.08^{+0.03}_{-0.03}$    &  $0.56^{+0.03}_{-0.03}$  & $0.31^{+0.03}_{-0.03}$      \\
SDSS J140025.40+314910.6 & 20.28 & $-26.83$ & 1.15 &$2.23_{-0.59}^{+0.71}$   & $1.77 ^{+0.57}_{-0.47}$   & $11.74 ^{+3.76}_{-3.12}$  & $1.76^{+0.76}_{-0.37}$ & 0.35 & 31.13 & $-0.76$ & 34.05 & 2.81 & $-1.33^{+0.05}_{-0.05}$    &  $0.32^{+0.05}_{-0.05}$  & $0.12^{+0.05}_{-0.05}$      \\
SDSS J154824.01+333500.1 & 20.35 & $-26.76$ & 2.40 & $10.66_{-1.41}^{+1.43}$ & $10.52^{+1.28}_{-1.27}$   &$146.46^{+17.85}_{-17.62}$ & $2.49^{+0.41}_{-0.21}$ & 0.32 & 31.10 & $-0.71$ & 34.31 & 3.08 & $-0.90^{+0.02}_{-0.02}$    &  $0.75^{+0.02}_{-0.02}$  & $0.51^{+0.02}_{-0.02}$        \\
\hline
\end{tabular}
\end{threeparttable}
\end{table}

\newcolumntype{H}{>{\setbox0=\hbox\bgroup}c<{\egroup}@{}}

\begin{table}
\centering
\caption{X-ray, optical/UV, and radio properties of $z>5.5$ RLQs. Objects are sorted by ascending RA. (2) AB magnitude at observed frame $1450(1+z)$ \si{\angstrom}, extrapolated from the flux density at rest-frame 2500 \angstrom assuming $\alpha_\nu=-0.5$. (3) Absolute magnitude at rest-frame 1450 \AA. (7) Flux density at $2/(1+z)$ keV in units of $10^{-31}$ \si{erg.cm^{-2}.s^{-1}.Hz^{-1}}. The remaining columns are identical to those in Table \ref{tab:lowzresults}. }
\label{tab:highzresults}
\begin{threeparttable}[b]
\begin{tabular}{lcccccccccccccHcc}
\hline
   Object name &  $m_\mathrm{1450}$ & $M_\mathrm{1450}$ & $N_\mathrm{H}$ &C.R. &  $F_\mathrm{X}$ &  $f_\mathrm{2keV}$ &    $\Gamma_\mathrm{X}$ & $f_\mathrm{2500\angstrom}$ & $\log L_\mathrm{2500\angstrom}$ &  $\alpha_\mathrm{r}$ &   $\log L_\mathrm{r}$ & $\log R$ &  $\alpha_\mathrm{ox}$& $\tilde{\alpha_\mathrm{ox}}$  &  $\Delta\alpha_\mathrm{ox,RQQ}$ &  $\Delta\alpha_\mathrm{ox,RLQ}$ \\
   (1) & (2) & (3) & (4) & (5) & (6) & (7) & (8) & (9) & (10) & (11) & (12) & (13) &(14) & (15) & (15) & (16) \\
   \hline
   \multicolumn{17}{c}{\textit{Chandra} Cycle 23 Obejcts}\\
PSO J055.4244$-$00.8035 & 20.70 & $-25.91$ &  5.89 & $0.64^{+0.24}_{-0.19}$  &  $1.02^{+0.38}_{-0.31}$ & $1.71^{+0.64}_{-0.52}$ &    $2.45^{+1.20}_{-0.51}$ & 0.25 &  31.12  &  $-0.99$       &  33.19 &  1.94 &  $-1.22^{+0.05}_{-0.06}$ & $-1.27$ &  $0.44^{+0.05}_{-0.06}$ &   $0.30^{+0.05}_{-0.06}$ \\
PSO J135.3860+16.2518 & 21.04 & $-25.55$   &  3.50 & $0.50^{+0.22}_{-0.17}$  &  $0.92^{+0.41}_{-0.32}$ & $2.71^{+1.20}_{-0.95}$ &    $2.97_{-0.51}^{+3.77}$ & 0.18 &  30.98  & $-0.55$ &  33.46 &  2.36 &  $-1.09^{+0.06}_{-0.07}$ & $-1.27$ &  $0.55^{+0.06}_{-0.07}$ &   $0.38^{+0.06}_{-0.07}$\\
 \multicolumn{17}{c}{Archival Objects}\\
SDSS J083643.85+005453.3 &18.77& $-27.07$ &  4.81 & - & $1.40^{+0.33}_{-0.28}$ & $1.61^{+37}_{-0.33}$ & $2.10^{+0.87}_{-0.31}$ & $1.48$ &  $31.91$ & $-0.55$   & 33.07 &  1.21 &  $-1.52^{+0.03}_{-0.04}$ & $-1.42$ &  $0.24^{+0.03}_{-0.04}$ &  $0.16^{+0.03}_{-0.04}$\\
PSO J172.3556+18.7734 & 21.68 & $-25.22$  &  3.48 & - &$0.24^{+0.09\text{  a}}_{-0.08} $& $0.28^{+0.10}_{-0.10}$ &  - & $0.10$ &  $30.85$ & $-1.31$   & 33.04 &  2.07 & $-1.37^{+0.05}_{-0.07}$ & & $0.24^{+0.05}_{-0.07}$ & $0.10^{+0.05}_{-0.07}$  \\
CFHQS J142952+544717     &21.03& $-25.71$ &  1.15 & $2.08^{+0.25}_{-0.25}$ & $2.72^{+0.33}_{-0.33}$ & $3.23^{+0.39}_{-0.39}$ & $2.08^{+0.33}_{-0.19}$ & $0.19$ &  $31.04$ & $-0.54$   & 33.40 &  2.24 &  $-1.06^{+0.02}_{-0.02}$ & $-1.01$ &  $0.58^{+0.02}_{-0.02}$ &  $0.42^{+0.02}_{-0.02}$ \\
VIKING J231818.35$-$311346.3 & 21.32 & $-25.49$ & 1.10 & $0.09^{+0.04}_{-0.03}$ & $0.13^{+0.06}_{-0.05}$ & $0.16^{+0.07}_{-0.06}$ & $>2.12$ & 0.14 & 30.95 & $-0.80$ & 32.76 & 1.69 & $-1.51^{+0.06}_{-0.08}$ & & $0.12^{+0.06}_{-0.08}$ & $0.00^{+0.06}_{-0.08}$\\
PSO J352.4034$-$15.3373  &21.36& $-25.29$ &  1.67 & - & $0.34^{+0.04}_{-0.04}$ & $0.43^{+0.05}_{-0.05}$ &$2.19_{-0.19}^{+0.30}$ &  
0.14& $30.89$ & $-0.95$   & 34.18 &  3.18 & $-1.35^{+0.02}_{-0.02}$ & & $0.27^{+0.02}_{-0.02}$ &  $0.03^{+0.02}_{-0.02}$\\
 \hline
 
\end{tabular}
\begin{tablenotes}
    \item[a] Flux in the observed frame 0.5--2 keV is calculated from the hard band (1--4.5 keV) count rate assuming $\Gamma_\mathrm{X}=2.0$.
\end{tablenotes}
\end{threeparttable}
\end{table}
\end{landscape}

\newcolumntype{H}{>{\setbox0=\hbox\bgroup}c<{\egroup}@{}}

\begin{landscape}
\begin{table}
\centering
\caption{Multiwavelength properties of the 41 HRLQs at $4<z<5.5$ (\citetalias{wu13}; \citetalias{zhu19}; this work) and 7 RLQs at $z>5.5$. (1) Object name. (2) Flux density at observed-frame 150 MHz, taken from the TIFR GMRT Sky Survey alternative data release \citep[TGSS ADR;][]{Intema2017}{}{}. (3) Flux density at observed-frame 326 MHz, taken from the Westerbork Northern Sky Survey \citep[WENSS][]{wenss}{}{}. (4) Flux density at observed-frame 366 MHz, taken from the Texas Survey of Radio Sources \citep[TXS;][]{txs}{}{}. (5) Flux density at observed-frame 1.4 GHz, taken from FIRST or NVSS. (6) Flux density at observed-frame 3GHz, taken from the VLASS Epoch 1 Quick Look Catalog \citep[][]{Gordon2021}{}{} (7) Flux density at observed-frame 5 GHz, taken from the Green Bank 6-cm survey \citep[GB6;][]{GB6}{}{} unless otherwise specified. (8) Radio spectral index between observed-rame 1.4 GHz and a lower frequency. (9) Radio spectral index between observed-rame 1.4 GHz and a higher frequency. (10) Radio spectral index at observed-frame \SI{1.4}{GHz}, taking the average of $\alpha_\mathrm{low}$ and $\alpha_\mathrm{high}$. (11) VLBI observed band that is closest to 1.4 GHz. (12) VLBI flux density observed at the band in (11). (13) The flux density at corresponding VLBI bands inferred from $f_\mathrm{1.4GHz}$ assuming radio spectral slope to be $\alpha_\mathrm{high}$. (14) Angular size of the compact component of the VLBI image. If the best-fit Gaussian component is elliptical, we adopt $\theta_\mathrm{VLBI}=\sqrt{\theta_\mathrm{maj}\theta_\mathrm{min}}$, where $\theta_\mathrm{maj}$ and $\theta_\mathrm{min}$ are angular sizes of the major and minor axes, respectively. $\theta_\mathrm{VLBI}$ is not provided if the source does not have available size estimation (observed by astrometric calibration surveys or not explicitly reported by authors) or extended structures have significant contributions to the total flux.  (15) VLBI references: (a) \citet{Coppejans2016}, (b) \citet{petrov2013}, (c) \citet{gordon2016}, (d) \citet{krezinger2022}, (e) \citet{petrov2021}, (f) \citet{frey2010}, (g) \citet{frey2018}, (h) \citet{momjian2004}, (i) \citet{frey2015}, (j) \citet{hunt2021},  (k) \citet{cao2017}, (l) \citet{Pushkarev2012}, (m) \citet{frey2003}, (n) \citet{momjian2021} (o) \citet{frey2011}, (p) \citet{zhang2022}, (q) \citet{Momjian2018}, (r) Astrogeo: \href{http://astrogeo.org}{http://astrogeo.org}.} 
\label{tab:radioprop}
\begin{tabular}{lHcccccccccHHHHHccccc}
\hline
\hline
Object name& $z$ &$f_\mathrm{150MHz}$ & $f_\mathrm{326MHz}$ & $f_\mathrm{366MHz}$ & $f_\mathrm{1.4GHz}$ & $f_\mathrm{3GHz}$ & $f_\mathrm{5GHz}$  & $\alpha_\mathrm{low}$ & $\alpha_\mathrm{high}$ & $\alpha_\mathrm{r}$& $\log R$ & $\Gamma_\mathrm{X}$ & $\alpha_\mathrm{ox}$ & $\Delta\alpha_\mathrm{ox,RQQ}$ & $\Delta\alpha_\mathrm{ox,RLQ}$ & VLBI band & $f_\mathrm{VLBI}$ &  $f_\mathrm{lowres}$ &$\theta_\mathrm{VLBI}$ & Ref. \\
           &     &    (mJy)           &        (mJy)      &    (mJy)          &     (mJy)           &      (mJy)        &       (mJy)       &                     &                         &                         &                    &                      &                                &                               &        &    (GHz)        &   (mJy)   &   (mJy)& (mas) & \\
    (1)    &     &     (2)            &          (3)      &       (4)         &      (5)            &       (6)         &        (7)        &         (8)             &                 (9)            &       (10)                  &            &              &                        &                      &    & (11) & (12) & (13)&(14) & (15)\\
\hline

\multicolumn{20}{c}{HRLQs at $4<z<5.5$ from this paper}\\
         SDSS J0825+1234 & 4.377 &       -             &     -              &       - &          $  16.68  \pm  0.15  $ &     $  8.96 \pm 0.12   $ &                -          &            -   & $ -0.82 $ & $ -0.82 $  & 2.82  &  $ 1.09^{+1.47}_{-0.52} $ & $ -1.69 $ & $ -0.08 $ &  $ -0.27 $  &  -    &    - &        -       & - &            - \\
         SDSS J0940+0526 & 4.503 &  $ 432.7 \pm44.1 $  &     -              &       - &          $  55.66  \pm  0.14  $ &     $ 37.68 \pm 0.14   $ &                -          &  $     -0.92 $ & $ -0.51 $ & $ -0.72 $  & 3.32  &  $ 2.21^{+0.58}_{-0.30} $ & $ -1.07 $ & $  0.55 $ &  $  0.31 $  &   1.7 & 18.3 &     50.4       & 1.91 &  (a)  \\
         SDSS J1047+0947 & 4.233 &  $ 21.0 \pm 3.3  $  &     -              &       - &          $  18.89  \pm  0.15  $ &     $ 25.47 \pm 0.12   $ &                -          &  $     -0.05 $ & $  0.39 $ & $ 0.17  $  & 2.59  &         -                 &    -      &     -     &     -       &   -   &   -  &       -        & - &          -\\
         SDSS J1156+4443 & 4.310 &  $ 282.9 \pm28.6 $  &  $ 220.0 \pm 4.0 $ &       - &          $  66.23  \pm  0.14  $ &     $ 35.13 \pm 0.14   $ &                -          &  $     -0.82 $ & $ -0.83 $ & $ -0.83 $  & 3.59  &  $ >1.66$                 & $ -1.46 $ & $  0.14 $ &  $ -0.12 $  &  -    &    - &        -       & - &            - \\
         SDSS J1253+5248 & 4.115 &  $  91.9 \pm10.5 $  &  $  89.0 \pm 3.3 $ &       - &          $  55.92  \pm  0.15  $ &     $ 40.52 \pm 0.18   $ &                -          &  $     -0.32 $ & $ -0.42 $ & $ -0.37 $  & 3.58  &  $ 1.52^{+1.40}_{-0.43} $ & $ -1.39 $ & $  0.18 $ &  $ -0.07 $  &  -    &    - &        -       & - &            - \\
         SDSS J1400+3149 & 4.690 &       -             &     -              &       - &          $  20.16  \pm  0.14  $ &     $ 11.33 \pm 0.13   $ &                -          &            -   & $ -0.76 $ & $ -0.76 $  & 2.81  &  $ 1.76^{+0.76}_{-0.38} $ & $ -1.33 $ & $  0.31 $ &  $  0.12 $  &   1.7 & 10.7 &     17.4       & 3.01 &  (a) \\
         SDSS J1538+4244 & 4.099 &       -             &     -              &       - &          $  11.71  \pm  0.13  $ &     $  6.80 \pm 0.13   $ &                -          &            -   & $ -0.71 $ & $ -0.71 $  & 2.71  &       -                   & $ -1.56 $ & $ <0.04 $ &  $ <-0.14$  &  -    &    - &        -       & - &            -   \\
         SDSS J1548+3335 & 4.678 &    $ 135.1\pm 14.9   $ &       $   107.0 \pm 3.6    $ &       - &          $  37.62   \pm 0.14  $ &     $ 22.01 \pm 0.11   $ &                -          &  $     -0.72 $ & $ -0.70 $ & $ -0.71 $  & 3.08  &  $ 2.49^{+0.41}_{-0.21} $ & $ -0.90 $ & $  0.73 $ &  $  0.51 $ & 1.7 & 10.0 & 32.8 & 1.76 & (a)   \\
         SDSS J1655+2834 & 4.048 &        -               &            -                 &       - &          $  23.01   \pm 0.14  $ &     $ 23.63 \pm 0.11   $ &         $  22 \pm 4 $     &            -   & $ -0.04 $ & $ -0.04 $  & 2.75  &  $ 1.42^{+0.92}_{-0.40} $ & $ -1.41 $ & $  0.21 $ &  $  0.02 $ &     - & -  &  -   & - &   -    \\

\multicolumn{20}{c}{HRLQs at $4<z<5.5$ from \citetalias{wu13} and \citetalias{zhu19}}\\
           PSS 0121+0347 & 4.130  &            -        &           -       &          -      & $ 72.95 \pm 0.20  $ & $ 58.11\pm0.17 $ &   $  51   \pm  7    $  &         -    & $ -0.28 $       &  $ -0.28 $ & 2.57 & $ 2.10^{+0.40}_{-0.30} $ & $ -1.47 $  & $  0.28 $   &   $  0.04 $ &   2.2 &  65.0 &    64.3           & - &    (b) \\  
             \textbf{B3 0254+434} & 4.067  & $   72.2 \pm 7.6$   &  $ 108.0\pm 3.9 $ &          -      & $148.10  \pm  5.30 $& $325.16\pm1.42 $ &   $ 159   \pm  14   $  &   $  0.22 $  &$1.03^\mathrm{a}$&  $  0.63 $ & 3.29 & $ 1.44^{+0.21}_{-0.11} $ & $ -1.23 $  & $  0.44 $   &   $  0.15 $ &   2.3 & 353.0 &   246.8  & - &    (c)  \\  
         SDSS J0304+0046 & 4.266  & $   36.2 \pm 5.7$   &        -          &          -      & $ 20.64  \pm  0.11 $& $ 16.33\pm0.09 $ &           -            &   $ -0.25 $  & $ -0.31 $       &  $ -0.28 $ & 3.15 & $ >1.79 $                & $ -1.23 $  & $  0.35 $   &   $  0.11 $ &   1.7 &  15.0 &    19.4           & 4.78 &    (d) \\  
          \textbf{PMN J0324$-$2918} & 4.630  & $   51.7 \pm 6.8$   &        -          &          -      & $236.50  \pm  7.10 $& $177.33\pm1.09 $ &           -            &   $  0.68 $  & $ -0.38 $       &  $  0.15 $ & 2.95 & $ 1.80^{+0.40}_{-0.40} $ & $ -1.40 $  & $  0.35 $   &   $  0.08 $ &   2.3 & 238.0 &  199.2 & - &   (a) \\  
          PMN J0525$-$3343 & 4.401  & $   73.5 \pm 11.7$   &        -          &          -      & $188.30  \pm  5.70 $& $107.93\pm1.01 $ &           -            &   $  0.42 $  & $ -0.73 $       &  $ -0.15 $ & 2.90 & $ 1.67^{+0.02}_{-0.02} $ & $ -1.17 $  & $  0.58 $   &   $  0.31 $ &   4.3 &  71.0 &    83.0        & - &   (e) \\  
         SDSS J0813+3508 & 4.929  & $  175.1 \pm18.8$   &  $ 153.0\pm3.8  $ &          -      & $ 48.99  \pm  0.16 $& $ 23.94\pm0.14 $ &           -            &   $ -0.78 $  & $ -0.94 $       &  $ -0.86 $ & 2.61 & $ 1.35^{+0.44}_{-0.17} $ & $ -1.55 $  & $  0.16 $   &   $ -0.07 $ &   1.6 &  17.1 &    43.2           & 1.18 &     (f) \\  
         SDSS J0835+1825 & 4.412  & $   72.8 \pm 7.9$   &       -           &          -      & $ 51.49  \pm  0.13 $& $ 56.16\pm0.13 $ &   $  40  \pm     5  $  &   $ -0.16 $  & $ -0.20 $       &  $ -0.18 $ & 3.19 & $ 1.56^{+0.15}_{-0.11} $ & $ -1.12 $  & $  0.51 $   &   $  0.25 $ &     - &     - &    -              & - &             - \\  
              Q0906+6930 & 5.480  & -                   &  $  35.0\pm3.9  $ &          -      & $ 92.80  \pm  2.80 $& $106.27\pm0.66 $ &   $ 106  \pm    10  $  &   $  0.67 $  & $  0.10 $       &  $  0.39 $ & 3.01 & $ 1.60^{+0.10}_{-0.10} $ & $ -1.31 $  & $  0.40 $   &   $  0.13 $ &   2.3 & 118.3 &    97.5           & 0.98 &    (g) \\  
         SDSS J0913+5919 & 5.122  &        -            &      -            &          -      & $ 17.45  \pm  0.16 $& $  9.40\pm0.11 $ &           -            &         -    & $ -0.81 $       &  $ -0.81 $ & 2.72 & $ >0.71 $                & $ -1.76 $  & $ -0.09 $   &   $ -0.32 $ &    1.4&  19.0 &    17.4           & 1.3 &    (h) \\  
         SDSS J1021+2209 & 4.262  & $  444.5 \pm44.7$   &      -            &          -      & $134.67  \pm  0.47 $& $127.64\pm0.20 $ &   $ 108   \pm   10  $  &   $ -0.53 $  & $ -0.17 $       &  $ -0.35 $ & 3.75 & $ >1.30 $                & $ -1.20 $  & $  0.41 $   &   $  0.10 $ &    4.3&  97.0 &   111.2           & - &    (e) \\  
         SDSS J1026+2542 & 5.304  & $  422.9 \pm42.4$   &      -            &  $ 406\pm  24  $& $230.81  \pm  0.14 $& $143.32\pm0.18 $ &   $ 142   \pm   13  $  &   $ -0.42 $  & $ -0.38 $       &  $ -0.40 $ & 3.54 & $ 1.73^{+0.32}_{-0.31} $ & $ -1.31 $  & $  0.39 $   &   $  0.07 $ &   1.7 & 137.5 &   213.6           & - &      (i) \\  
           \textbf{RX J1028$-$0844} & 4.276  & $  127.4 \pm13.3$   &      -            &          -      & $269.20  \pm  8.10 $& $124.38\pm0.16 $ &           -            &   $  0.33 $  & $ -1.01 $       &  $ -0.34 $ & 3.33 & $ 1.40^{+0.03}_{-0.03} $ & $ -1.09 $  & $  0.63 $   &   $  0.34 $ &   2.3 & 147.0&   163.0 & - &      (j) \\  
         SDSS J1113+4645 & 4.468  & $   34.0 \pm 4.2$   &      -            &          -      & $ 24.91  \pm  0.15 $& $ 22.13\pm0.12 $ &   $  20   \pm    4  $  &   $ -0.14 $  & $ -0.17 $       &  $ -0.16 $ & 2.81 & $ 1.51^{+0.59}_{-0.21} $ & $ -1.47 $  & $  0.11 $   &   $ -0.05 $ &     - &     - &    -              & - &             - \\  
          PMN J1155$-$3107 & 4.300  &        -            &      -            &          -      & $ 82.00  \pm  2.50 $& $163.71\pm1.48 $ &           -            &         -    & $  0.91 $       &  $  0.91 $ & 2.73 & $ >1.33 $                & $ -1.36 $  & $  0.36 $   &   $  0.12 $ &   2.3 & 129.9 &   128.8         & - & (c) \\  
         SDSS J1235$-$0003 & 4.673  &        -            &      -            &          -      & $ 18.27  \pm  0.15 $& $ 11.61\pm0.15 $ &           -            &         -    & $ -0.59 $       &  $ -0.59 $ & 3.05 &      -                   & $ -1.22 $  & $  0.39 $   &   $  0.16 $ &    1.4&  18.8 &    18.3         & $<3.3$ & (h) \\  
         SDSS J1237+6517 & 4.301  & $  115.5 \pm11.7$   &  $ 81.0\pm3.3  $  &          -      & $ 25.50  \pm  0.90 $& $ 15.70\pm0.24 $ &           -            &   $ -0.79 $  & $ -0.64 $       &  $ -0.71 $ & 2.50 & $ 1.92^{+0.90}_{-0.30} $ & $ -1.52 $  & $  0.16 $   &   $ -0.05 $ &     - &     - &    -              & - &             - \\  
         \textbf{SDSS J1242+5422} & 4.750  & $   19.7 \pm 3.1$   &  $ 30.0\pm2.8  $  &          -      & $ 19.74  \pm  0.13 $& $ 10.41\pm0.12 $ &           -            &   $ -0.29 $  & $ -0.84 $       &  $ -0.56 $ & 2.67 & $ 2.38^{+1.13}_{-0.33} $ & $ -1.24 $  & $  0.43 $   &   $  0.21 $ &  1.6  &  17.7 &    17.7  & 0.67 &      (f) \\  
        CLASS J1325+1123 & 4.415  &        -            &      -            &          -      & $ 69.39  \pm  0.14 $& $ 53.02\pm0.14 $ &   $  62   \pm     8 $  &         -    & $ -0.09 $       &  $ -0.09 $ & 2.72 & $ 1.80^{+0.50}_{-0.40} $ & $ -1.53 $  & $  0.19 $   &   $ -0.05 $ &    1.7&  62.7 &    68.2           & 0.76 &  (d) \\  
         SDSS J1348+1935 & 4.404  & $   52.1 \pm 6.9$   &      -            &          -      & $ 49.25  \pm  0.15 $& $ 42.75\pm0.13 $ &   $  38   \pm     5 $  &   $ -0.03 $  & $ -0.20 $       &  $ -0.11 $ & 2.97 & $ 1.56^{+0.34}_{-0.17} $ & $ -1.37 $  & $  0.29 $   &   $  0.04 $ &     - &     - &    -              & - &             - \\  
         SDSS J1412+0624 & 4.467  & $  122.8 \pm16.2$   &      -            &          -      & $ 43.04  \pm  0.14 $& $ 27.97\pm0.15 $ &   $  34   \pm     6 $  &   $ -0.47 $  & $ -0.19 $       &  $ -0.33 $ & 2.70 & $ 1.33^{+0.43}_{-0.38} $ & $ -1.51 $  & $  0.18 $   &   $ -0.06 $ &    1.7&  18.8 &    41.5           & $<2.66$ & (d) \\  
         SDSS J1420+1205 & 4.027  & $  403.7 \pm40.5$   &      -            &  $ 248 \pm 29 $ & $ 83.77  \pm  0.14 $& $ 48.41\pm0.45 $ &   $  47   \pm     6 $  &   $ -0.81 $  & $ -0.45 $       &  $ -0.63 $ & 3.05 & $ 1.61^{+0.35}_{-0.32} $ & $ -1.34 $  & $  0.33 $   &   $  0.06 $ &    1.7&  47.9 &    76.8           & - &        (k) \\  
            \textbf{GB 1428+4217} & 4.715  & $  174.3 \pm18.0$   &  $ 238.0\pm3.2  $ &  $ 222 \pm 18 $ & $211.27  \pm  0.15 $& $125.79\pm0.13 $ &   $ 337   \pm    30 $  &   $ -0.08 $  & $  0.37 $       &  $  0.14 $ & 3.06 & $ 1.73^{+0.03}_{-0.03} $ & $ -0.93 $  & $  0.80 $   &   $  0.52 $ &   2.3 & 150.0 &   253.9  & - &   (l) \\  
            \textbf{GB 1508+5714} & 4.313  & $  202.2 \pm20.7$   &  $ 218.0\pm3.6  $ &  $ 222 \pm 15 $ & $248.07  \pm  0.14 $& $243.22\pm1.44 $ &   $ 292   \pm    26 $  &   $  0.09 $  & $  0.13 $       &  $  0.11 $ & 3.87 & $ 1.55^{+0.05}_{-0.05} $ & $ -0.96 $  & $  0.67 $   &   $  0.34 $ &   2.3 & 276.6 &   264.6  & - &   (l) \\  
         SDSS J1535+0254 & 4.388  & $  150.1 \pm15.3$   &       -           &          -      & $ 78.82  \pm  0.15 $& $ 46.13\pm0.16 $ &   $  53   \pm     7 $  &   $ -0.29 $  & $ -0.31 $       &  $ -0.30 $ & 3.14 & $ 1.32^{+0.10}_{-0.09} $ & $ -1.12 $  & $  0.55 $   &   $  0.28 $ &   5   &  68.0 &    53.1           & - &   (r)\\ 
         SDSS J1605+2728 & 4.024  & $   43.6 \pm 4.9$   &       -           &          -      & $ 11.66  \pm  0.14 $& $  6.16\pm0.23 $ &           -            &   $ -0.59 $  & $ -0.84 $       &  $ -0.71 $ & 2.61 &     -                    & $ -1.31 $  & $  0.30 $   &   $  0.10 $ &     - &     - &    -              & - &             - \\  
         SDSS J1612+4702 & 4.350  & $  139.2 \pm14.6$   &  $ 118.0\pm3.6  $ &          -      & $ 52.26  \pm  0.15 $& $ 44.01\pm0.13 $ &   $  30   \pm    4  $  &   $ -0.56 $  & $ -0.44 $       &  $ -0.50 $ & 3.07 & $ >1.50 $                & $ -1.53 $  & $  0.13 $   &   $ -0.13 $ &     - &     - &    -              & - &             - \\

\hline
\end{tabular}
\end{table}
\end{landscape}

\addtocounter{table}{-1}

\begin{landscape}
\begin{table}
\centering
\caption{Continued} 
\begin{threeparttable}[b]
\begin{tabular}{lHcccccccccHHHHHccccc}
\hline
\hline
Object name& $z$ &$f_\mathrm{150MHz}$ & $f_\mathrm{326MHz}$ & $f_\mathrm{366MHz}$ & $f_\mathrm{1.4GHz}$ & $f_\mathrm{3GHz}$ & $f_\mathrm{5GHz}$  & $\alpha_\mathrm{low}$ & $\alpha_\mathrm{high}$ & $\alpha_\mathrm{r}$& $\log R$ & $\Gamma_\mathrm{X}$ & $\alpha_\mathrm{ox}$ & $\Delta\alpha_\mathrm{ox,RQQ}$ & $\Delta\alpha_\mathrm{ox,RLQ}$ & VLBI band & $f_\mathrm{VLBI}$ & $f_\mathrm{lowres}$ & $\theta_\mathrm{VLBI}$ & Ref. \\
           &     &    (mJy)           &        (mJy)      &    (mJy)          &     (mJy)           &      (mJy)        &       (mJy)       &                     &                         &                         &                    &                      &                                &                               &        &    (GHz)        &   (mJy)   &   (mJy)&(mas)&  \\
    (1)    &     &     (2)            &          (3)      &       (4)         &      (5)            &       (6)         &        (7)        &         (8)             &                 (9)            &       (10)                  &            &              &                        &                      &    & (11) & (12) & (13)&(14) &(15)\\
\hline
            \textbf{SDSS J1659+2101} & 4.784  & $   27.1 \pm 4.0$   &          -        &          -      & $ 28.73  \pm  0.15 $& $ 23.43\pm1.10 $ &           -            &   $  0.03 $  & $ -0.27 $       &  $ -0.12 $ & 2.56 & $ 1.75^{+0.40}_{-0.36} $ & $ -1.39 $  & $  0.30 $   &   $  0.07 $ &   1.6 &  29.3 &    27.7 & 3.07  &        (f) \\  
             \textbf{GB 1713+2148} & 4.011  & $ 1080.6 \pm108.1$  &          -        & $1118 \pm50    $& $446.40  \pm14.90  $& $248.42\pm1.58 $ &   $ 306    \pm 27   $  &   $ -0.68 $  & $ -0.30 $       &  $ -0.49 $ & 4.50 & $ 1.63^{+0.24}_{-0.23} $ & $ -1.16 $  & $  0.42 $   &   $  0.05 $ &  4.3  &  55.0 &   318.8   & -  &      (e) \\  
             PMN J1951+0134 & 4.114  & $   47.4 \pm  7.8$  &          -        &          -      & $ 99.10  \pm 3.00  $& $254.90\pm0.22 $ &   $ 151    \pm 14   $  &   $  0.33 $  &$1.24^\mathrm{a}$&  $  0.79 $ & 3.04 & $ 1.54^{+0.33}_{-0.30} $ & $ -1.23 $  & $  0.45 $   &   $  0.20 $ &  2.3  & 259.0 &   183.4          & -  &      (c) \\  
            \textbf{PMN J2134$-$0419} & 4.346  & $  554.4 \pm 56.2$  &          -        &  $ 622 \pm 58  $& $303.79  \pm 0.15  $& $313.74\pm2.09 $ &                  -     &   $ -0.53 $  & $  0.04 $       &  $ -0.25 $ & 3.40 & $ 1.70^{+0.38}_{-0.18} $ & $ -1.38 $  & $  0.33 $   &   $  0.02 $ &   1.7 & 236.6  & 306.2 & 1.42  &       (k)\\  
          SDSS J2220+0025 & 4.220  & $  644.7 \pm 64.5$  &          -        &          -      & $ 92.65  \pm 0.10  $& $ 40.24\pm0.45 $ &   $  37    \pm  7  $   &   $ -0.87 $  & $ -0.72 $       &  $ -0.79 $ & 3.26 & $ >1.76 $                & $ -1.29 $  & $  0.34 $   &   $  0.07 $ &   1.7 &   8.2 &    80.6            & -  &       (k) \\  
         \textbf{ PMN J2314+0201} & 4.110  & $  103.8 \pm 11.7$  &          -        &          -      & $117.83  \pm 0.12  $& $ 87.81\pm0.14 $ &   $  84    \pm  9  $   &   $  0.06 $  & $ -0.27 $       &  $ -0.10 $ & 3.13 & $ 1.33^{+0.33}_{-0.15} $ & $ -1.47 $  & $  0.21 $   &   $ -0.06 $ &  5    &  77.6 &    83.6    & -  &      (r) \\

\multicolumn{20}{c}{RLQs at $z>5.5$}\\

     PSO J055$-$00 & 5.68 &             -        &  - & - &  $  2.14 \pm  0.14  $ &  $1.01\pm 0.13^{\mathrm{b}}$& -                           &    -                  &   $-0.99$  &  $-0.99$             &  $1.94$  & $2.45^{+1.20}_{-0.51}$  & $-1.22$ & $ 0.42 $&$  0.30 $  & -    &  -&-                   & - & - \\    
 SDSS J0836+0054 & 5.82 &             -        &  - & - &  $  1.53 \pm  0.15  $ &         -                   & $0.58\pm0.06^{\mathrm{c}}$  & $-0.76^{\mathrm{d}}$  &   $-0.33$  &   $-0.55$            &  $1.21$  & $1.97^{+0.85}_{-0.32}$  & $-1.53$ & $ 0.19 $&$  0.12 $  &  1.6 & 1.1&1.5                  & - &  (m) \\   
     PSO J135+16 & 5.63 &             -        &  - & - &  $  3.99 \pm  0.15  $ &  $  2.62\pm0.21 $           & -                           &    -                  &   $-0.55 $ &   $-0.55$            &  $2.36$  & $2.97^{+3.77}_{-0.51}$  & $-1.09$ & $ 0.53 $&$  0.38 $  & -    & -&-                      & - & - \\   
     PSO J172+18 & 6.82 & $<8.5^\mathrm{e}$    &  - & - &  $ 1.02\pm 0.14$      &   -                         &  -                          &  -                    &   $-1.31^\mathrm{f}$  &   $-1.31$ &    -     &           -             &     -   &    -    &     -     &  1.5 & 0.50 & $0.51^\mathrm{g}$ & 5.9 &  (n)\\
CFHQS J1429+5447 & 6.18 &             -        &  - & - &  $  2.93 \pm  0.15  $ &  $  2.07\pm0.11 $           & -                           & $-0.62^{\mathrm{h}}$  &   $-0.45 $ &   $-0.54$            &  $2.24$  & $2.50^{+0.20}_{-0.20}$  & $-0.88$ & $ 0.75 $&$  0.60 $  & 1.6  & 3.3 & 2.8                & 2.63 &  (o)  \\      
VIK J2318$-$3113 & 6.44 &          -           &  - & - &  $   0.52\pm 0.17^{\mathrm{i}}   $ &  $  0.40\pm0.15^\mathrm{j}$            &  $0.104\pm 0.034^{\mathrm{k}}$            & $-0.44^{\mathrm{l}}$  &   $-1.16$  &   $-0.80$            &  -       &    -   &  -    &   -   &  -   & 1.57 & 0.48 & 0.44  & 2.3 &  (p) \\
     PSO J352$-$15 & 5.84 &    $163.1\pm20.7   $ &  - & - &  $ 14.90 \pm  0.70  $ &  $  7.89\pm0.15 $           &   -                         & $-1.07 $              &   $-0.83 $ &   $-0.95$            &  $3.18$  & $1.99^{+0.29}_{-0.28}$  & $-1.41$ & $ 0.20 $&$ -0.03 $  &  1.5 & 6.57 & 14.1            & - &  (q) \\   
        
\hline

\end{tabular}
\begin{tablenotes}
    \item[a] $f_\mathrm{3GHz}$ of these objects are substantially larger than $f_\mathrm{1.4GHz}$ and $f_\mathrm{5GHz}$. Use $f_\mathrm{3GHz}$ to calculate $\alpha_\mathrm{high}$ rather than the preferred $f_\mathrm{5GHz}$.
    \item[b] Forced photometry using {\sc aegean}.
    \item[c] $f_\mathrm{5GHz}$ of SDSS J083643.85+005453.3 is taken from \citet{petric2003}.
    \item[d] $\alpha_\mathrm{low}$ is calculated from $f_\mathrm{1.4GHz}$ and $f_\mathrm{323MHz}$ \citep[\SI{2.47}{mJy};][]{shao2020}{}{}.
    \item[e] A 3$\sigma$ upper limit obtained from \citet{banados2021}.
    \item[f] $\alpha_\mathrm{high}$ is calculated from flux densities at observed-frame 1.52 GHz and 2.87 GHz obtained from simultaneous follow-up radio observations reported by \citet{banados2021} to avoid possible variability.
    \item[g] $f\mathrm{lowres}$ is the flux density at observed-frame 1.52 GHz taken from \citet{banados2021} instead of extrapolating from FIRST observation at 1.4 GHz to avoid possible variability.
    \item[h] $\alpha_\mathrm{low}$ is calculated from $f_\mathrm{1.4GHz}$ and $f_\mathrm{144MHz}$ \citet[\SI{11.99}{mJy}; LoTSS DR2,][]{shimwell2022}{}{}. 
    \item[i] $f_\mathrm{1.4GHz}$ of VIK J2318$-$3113 is the flux density at observed-frame 1367 MHz taken from the Rapid ASKAP Continuum Survey \citep[RACS,][]{McConnell2020}{}{}.
    \item[j] $f_\mathrm{3GHz}$ of VIK J2318$-$3113 is taken from \citet{Ighina2022_VIK}, calculated from the VLASS Epoch 2 Quick Lood Image.
    \item[k] $f_\mathrm{5GHz}$ of VIK J2318$-$3113 is the flux density at observed-frame 5500 MHz taken from \citet{Ighina2022_VIK}.
    \item[l] $\alpha_\mathrm{low}$ is calculated from $f_\mathrm{399 MHz}$ \citep[0.89 mJy; EMU,][]{Norris2011}{}{} and $f_\mathrm{1.4GHz}$.
\end{tablenotes}
\end{threeparttable}
\end{table}
 \end{landscape}

\end{document}